\documentstyle[natbib209,2up,multicol,float]{arPrepn}
\twouparticle

\newcommand{\be}{\begin{equation} }
\newcommand{\ee}{\end{equation}}

\newcommand{\bea}{\begin{eqnarray}}
\newcommand{\eea}{\end{eqnarray}}
\newcommand{\vs}{\nonumber\\}
\newcommand{\like}{{\cal L}}

\newcommand{\secl}[1]{ \label{sec:#1}}
\newcommand{\secc}[1]{\S \ref{sec:#1}}
\newcommand{\eqnl}[1]{\label{eqn:#1}}
\newcommand{\eqnc}[1]{Equation (\ref{eqn:#1})}
\newcommand{\figl}[1]{\label{fig:#1}}
\newcommand{\figc}[1]{Figure \ref{fig:#1}}
\newcommand{\plal}[1]{\label{pla:#1}}
\newcommand{\plac}[1]{Plate \ref{pla:#1}}
\newcommand{\vc}[1]{{\bf #1}}
\newcommand{\tot}{{\rm tot}}
\newcommand{\ri}{{\rm ri}}
\newcommand{\ella}{\ell_{\rm a}}
\newcommand{\elld}{\ell_{\rm d}}
\newcommand{\kd}{k_{\rm d}}
\newcommand{\elleq}{\ell_{\rm eq}}

\newcommand{\apj}{Ap.\ J.}
\newcommand{\nat}{Nature}
\newcommand{\aap}{Astron. Astrophys.}
\newcommand{\aj}{Astron. J.}
\newcommand{\apjl}{Ap. J. Lett.}
\newcommand{\apjs}{Ap. J. Sup.}
\newcommand{\prd}{Phys. Rev. D.}
\newcommand{\prl}{Phys. Rev. Lett.}
\newcommand{\araa}{Annu. Rev. Astron. Astrophys.}
\newcommand{\mnras}{MNRAS}
\newcommand{\wm}{\Omega_m h^2}
\newcommand{\wb}{\Omega_b h^2}
\newcommand{\da}{D}
\newcommand{\curv}{\hat F}

\floatstyle{plain}
\newfloat{plate}{thp}{pla}
\floatname{plate}{Plate}

\renewcommand{\refname}{Literature Cited}
\def\spose#1{\hbox to 0pt{#1\hss}}
\def\simlt{\mathrel{\spose{\lower 3pt\hbox{$\mathchar"218$}}
     \raise 2.0pt\hbox{$\mathchar"13C$}}}
\def\simgt{\mathrel{\spose{\lower 3pt\hbox{$\mathchar"218$}}
     \raise 2.0pt\hbox{$\mathchar"13E$}}}
\def\simpropto{\mathrel{\spose{\lower 3pt\hbox{$\mathchar"218$}}
     \raise 2.0pt\hbox{$\propto$}}}

\begin{document}

\input epsf.tex

\input psfig.sty

\jname{Annu. Rev. Astron. and Astrophys.}
\jyear{2002}

\title{Cosmic Microwave Background Anisotropies}

\markboth{Hu \& Dodelson}{CMB Anisotropies}

\author{Wayne Hu$^{1,2,3}$ and Scott Dodelson$^{2,3}$ 
\affiliation{$^1$Center for Cosmological Physics, University of Chicago,
Chicago, IL 60637\\
$^2$NASA/Fermilab Astrophysics Center, P.O. Box 500, Batavia, IL 60510\\
$^3$Department of Astronomy and Astrophysics, University of Chicago, 
Chicago, IL 60637}}


\begin{abstract}
Cosmic microwave background (CMB) temperature 
anisotropies have and will continue to revolutionize 
our understanding of cosmology.  The recent discovery 
of the previously predicted acoustic peaks in the power spectrum 
has established a working cosmological model: a critical density 
universe consisting of mainly dark matter and dark energy,
which formed its structure through gravitational instability
from quantum fluctuations during an inflationary epoch.
Future observations should test this model and measure its 
key cosmological parameters with unprecedented precision.  
The phenomenology and cosmological 
implications of the acoustic peaks are developed in detail.  
Beyond the peaks, the yet to be detected
secondary anisotropies and polarization present opportunities 
to study the physics of inflation and the dark energy.  
The analysis techniques devised to extract cosmological information
from voluminous CMB data sets are outlined, given their 
increasing importance in experimental cosmology as a whole.
\end{abstract}

\maketitle

\section{INTRODUCTION}
\secl{intro}

The field of cosmic microwave background (CMB) anisotropies has 
dramatically advanced over the last decade (c.f.\ \citealt{WhiScoSil94}), 
especially on its observational front.
The observations have turned some of our boldest
speculations about our Universe into a working cosmological model:
namely, that the Universe is spatially flat, 
consists mainly of dark matter and dark energy, with the small amount of
ordinary matter necessary to explain the light element abundances, 
and all the rich structure in it formed through gravitational instability 
from quantum mechanical fluctuations when the Universe was a fraction 
of a second old.
Observations over the coming decade should pin
down certain key cosmological parameters with unprecedented 
accuracy \citep{Kno95,JunKamKosSpe96,BonEfsTeg97,ZalSpeSel97,EisHuTeg99}. 
These determinations
will have profound implications for astrophysics, as well as other
disciplines. Particle physicists, for example, will be able to study
neutrino masses, 
theories of inflation impossible to test at accelerators, and
the mysterious  dark energy or cosmological constant. 

For the twenty eight years between 
the discovery of the CMB \citep{PenWil65}
and the COBE DMR detection of $10^{-5}$ fluctuations in its temperature field
across the sky \citep{Smoetal92}, observers searched
for these anisotropies but found none except the 
dipole induced by our own motion \citep{SmoGorMul77}.
They learned the hard way that the CMB is remarkably
uniform. This is in stark contrast to the matter in the Universe, organized 
in very non-linear structures like galaxies and clusters. The disparity between
the smooth photon distribution and the clumpy matter distribution is 
due to radiation pressure.  Matter inhomogeneities 
grow due to gravitational instability, but pressure prevents 
the same process from occuring in the photons.
Thus, even though both inhomogeneities in the matter
in the Universe and anisotropies in the CMB apparently
originated from the same source, these appear very different today.

Since the photon distribution is very uniform, perturbations are
small, and linear response theory applies.  This is perhaps the
most important fact about CMB anisotropies. Since they are linear, predictions
can be made as precisely as their sources are specified.  
If the sources of the anisotropies are also linear
fluctuations, anisotropy 
formation falls in the domain of linear
perturbation theory.  There are then essentially
no phenomenological parameters that need to be introduced to account for
non-linearities or gas dynamics or any other of a host of astrophysical
processes that typically afflict cosmological observations.

CMB anisotropies in the working cosmological model, 
which we briefly review in \secc{observables}, fall almost entirely 
under linear perturbation theory.
The most important observables of the CMB are the power spectra 
of the temperature and polarization maps. 
Theory predicts, and now observations confirm, that the 
temperature power spectrum has a series of peaks and troughs. 
In \secc{acoustic}, we discuss the origin of these
acoustic peaks and their cosmological uses.
Although they are the most prominent features in the spectrum, 
and are the focus of the current generation of experiments,
future observations will turn to even finer
details, potentially revealing the physics 
at the two opposite ends of time. 
Some of these are discussed in \secc{beyond}.  
Finally, the past few years have witnessed important new
advances, introduced in \secc{data}, from a growing body of CMB data 
analysts on how best to extract the information 
contained in CMB data.  Some of the fruits of this labor have 
already spread to other fields of astronomy. 

\section{OBSERVABLES}
\secl{observables}

\subsection{Standard Cosmological Paradigm}
\secl{standard}

While a review of the standard cosmological paradigm is not
our intention (see \citealt{NarPad01} for a critical appraisal), 
we briefly introduce the observables necessary to parameterize it.

The expansion of the Universe is described by the scale
factor $a(t)$, set to unity today, and by the current expansion rate,
the Hubble constant $H_0=100h$ km sec$^{-1}$ Mpc$^{-1}$, with $h\simeq 0.7$
\citep{Freetal01}. 
The Universe is {\it flat} (no spatial curvature) if the total 
density is equal to the critical density,
$\rho_c=1.88h^2\times 10^{-29}$g cm$^{-3}$; it is {\it open} (negative
curvature) if the density is less than this and {\it closed} (positive
curvature) if greater.
The mean densities of different components of the 
Universe control $a(t)$ and are typically expressed today in units 
of the critical density $\Omega_i$, with an evolution with $a$ specified  
by equations of state $w_i=p_i/\rho_i$, where
$p_i$ is the pressure of the $i$th component. Density fluctuations
are determined by these parameters through the gravitational instability of
an initial spectrum of fluctuations.  

The working cosmological model contains photons, neutrinos, baryons,
cold dark matter and dark energy with densities proscribed within 
a relatively tight range.
For the radiation, 
$\Omega_r=4.17\times 10^{-5}h^{-2}$ ($w_r=1/3$). 
The photon contribution to the radiation is determined to high precision by the
measured CMB temperature, $T=2.728\pm0.004$K \citep{Fixetal96}.  The
neutrino contribution follows from the assumption of 3 neutrino species, 
a standard thermal history, and a negligible mass
$m_\nu \ll 1$eV.  Massive neutrinos have an equation of state $w_\nu = 1/3 \rightarrow 0$ as the particles become non-relativistic.  For $m_\nu \sim 1$eV this
occurs at $a \sim 10^{-3}$ and can leave a small but potentially measurable
effect on the CMB anisotropies \citep{MaBer95,DodGatSte96}.
 
For the ordinary matter or baryons, $\Omega_b\approx 0.02h^{-2}$ ($w_b\approx 0$) 
with statistical uncertainties at about the ten percent level determined through 
studies of the light element abundances 
(for reviews, see \citealt{BoeSte85,SchTur98,TytOmeSuzLub00}). 
This value is in strikingly good agreement with that implied by the
CMB anisotropies themselves as we shall see.
There is very strong evidence that there is also
substantial non-baryonic dark matter.
This dark matter must be close to cold ($w_m=0$) for the gravitational
instability paradigm to work \citep{Pee82} and
when added to the baryons gives
a total in non-relativistic matter of $\Omega_m\simeq 1/3$. 
Since the Universe appears to be flat, the total $\Omega_\tot$ must 
be equal to one. Thus, there
is a missing component to the inventory, dubbed {\it dark energy}, with
$\Omega_\Lambda\simeq 2/3$. 
The cosmological constant ($w_\Lambda=-1$)
is only one of several possible candidates but we will generally assume
this form unless otherwise specified. 
Measurements of an accelerated expansion from distant supernovae
\citep{Rieetal98,Peretal99} provide entirely independent evidence 
for dark energy in this amount.

The initial spectrum of density 
perturbations is assumed to be a power law with
a power law index or tilt of $n\approx 1$ corresponding
to a scale-invariant spectrum.
Likewise the initial spectrum of gravitational
waves is assumed to be scale-invariant, with an amplitude parameterized 
by the energy scale of inflation $E_i$, 
and constrained to be small compared with the initial density spectrum.  
Finally the formation of structure
will eventually reionize the Universe at some redshift $7 
\simlt z_{\rm ri} \simlt 20$.

Many of the features of the anisotropies 
will be produced even if these parameters fall outside the expected
range or even if the standard paradigm is incorrect.  Where appropriate, 
we will try to point these out.

\subsection{CMB Temperature Field}

The basic observable of the CMB is its intensity 
as a function of frequency and direction on the sky $\hat \vc{n}$.  
Since the CMB spectrum is an extremely good blackbody \citep{Fixetal96} with
a nearly constant temperature across the sky $T$, we generally
describe this observable in terms of a temperature fluctuation 
$\Theta(\hat \vc{n}) = \Delta T/T$.

If these fluctuations are Gaussian, then the multipole moments
of the temperature field
\begin{equation}
	\Theta_{\ell m}  = \int d{\hat \vc{n}} Y_{\ell m}^*(\hat \vc{n})
     \Theta(\hat \vc{n}) 
\eqnl{thdecompose}
\end{equation}
are fully characterized by their power spectrum
\begin{equation}
    \left< \Theta_{\ell m}^{*} \Theta_{\ell'm'} \right> =
   \delta_{\ell \ell'}\delta_{m m'} C_{\ell}\,,
\eqnl{cldef}
\end{equation}
whose values as a function of $\ell$ are independent in a given 
realization.
For this reason predictions and analyses are typically performed
in harmonic space.  On small sections of the sky where its curvature
can be neglected, the spherical harmonic analysis becomes ordinary
Fourier analysis in two dimensions. 
In this limit  
$\ell$ becomes the Fourier wavenumber. Since the
angular wavelength $\theta = 2\pi/\ell$, large multipole moments
corresponds to small angular scales with $\ell \sim 10^2$ representing
degree scale separations.  Likewise, since in this limit
the variance of the field is $\int d^2 \ell C_\ell/(2\pi)^2$, the
power spectrum is usually displayed as 
\begin{equation}
\Delta_T^2 \equiv {\ell(\ell+1) \over 2\pi} C_\ell T^2\,,
\eqnl{deltat}
\end{equation}
the power per logarithmic interval in wavenumber for $\ell \gg 1$.

\plac{cltt} (top) shows observations of $\Delta_T$ along with the prediction
of the working cosmological model,
complete with the acoustic peaks mentioned in \secc{intro}
and discussed extensively in \secc{acoustic}. 
While COBE first detected anisotropy on the
largest scales (inset), observations in the last decade 
have pushed the frontier to smaller and smaller scales 
(left to right in the figure). The MAP
satellite, launched in June 2001, will go out to $\ell\sim 1000$, while
the European satellite, Planck, scheduled for launch in 2007, will go 
a factor of two higher (see \plac{cltt} bottom). 

The power spectra shown in \plac{cltt} all begin at $\ell=2$ and
exhibit large errors at low multipoles.  The reason is that 
the predicted power spectrum is the
average power in the multipole moment $\ell$ an observer would see
in an ensemble of universes.  However a real observer is limited to
one Universe and one sky with its one set of $\Theta_{\ell m}$'s, 
$2\ell+1$ numbers for each $\ell$. 
This is particularly problematic
for the monopole and dipole ($\ell=0,1$).
If the monopole were larger in our
vicinity than its average value, we would have
no way of knowing it. 
Likewise for the dipole, we have no way of distinguishing a cosmological
dipole from our own peculiar motion with respect to the CMB rest frame.
Nonetheless, the monopole and dipole --
which we will often call simply $\Theta$ and $v_\gamma$ -- are of the utmost
significance in the early Universe. It is precisely the
spatial and temporal variation of these quantities, especially
the monopole,
which determines the
pattern of anisotropies we observe today. A distant observer sees
spatial variations in the local temperature or monopole, at a distance given by
the lookback time, as a fine-scale angular anisotropy.  Similarly,
local dipoles appear as a Doppler shifted temperature
which is viewed analogously.
In the jargon of the field, 
this simple projection is referred to as the {\it freestreaming} of 
power from the monopole and dipole to higher multipole moments.

\begin{plate}
\centerline{\epsfxsize=4.75in\epsffile{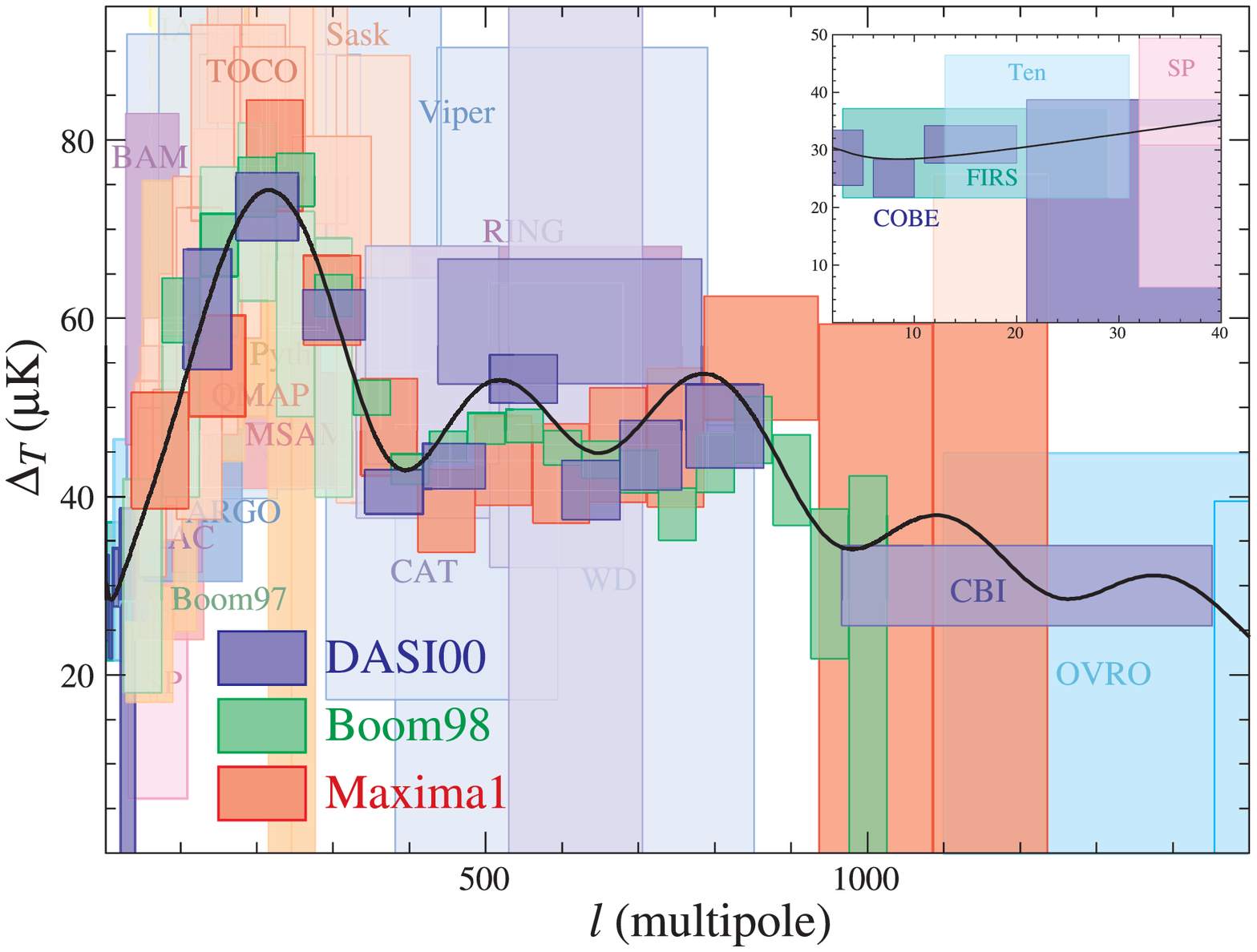}}
\centerline{\epsfxsize=4.75in\epsffile{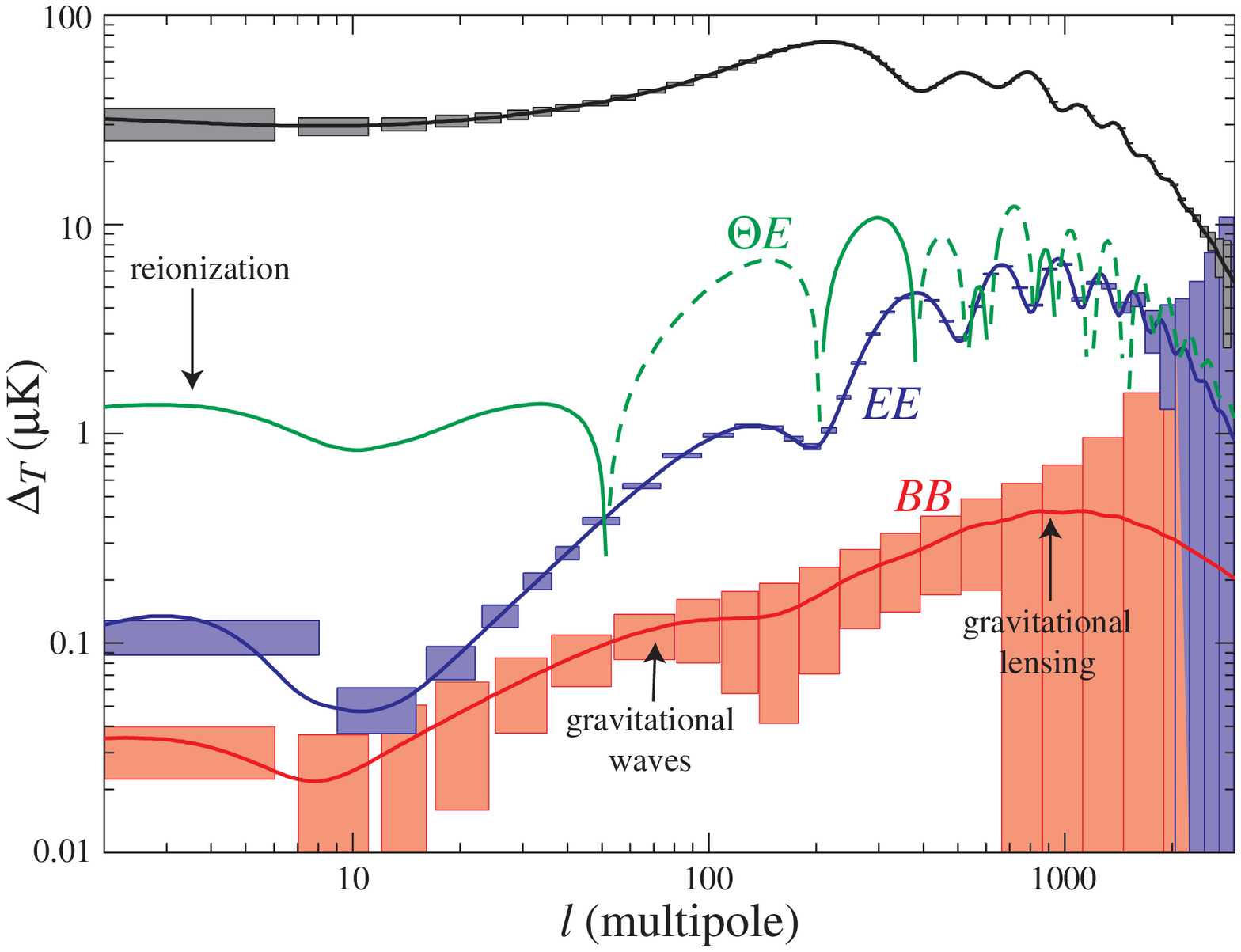}}
\caption{\footnotesize Top: temperature anisotropy data with boxes
representing $1$-$\sigma$ errors and approximate $\ell$-bandwidth. Bottom:
temperature and polarization spectra for $\Omega_{\rm tot}=1$,
$\Omega_\Lambda=2/3$, $\Omega_b h^2=0.02$, $\Omega_m h^2=0.16$, $n=1$, 
$z_{\rm ri}=7$, $E_i=2.2 \times 10^{16}$ GeV. 
Dashed lines represent negative cross correlation and boxes 
represent the statistical errors of the Planck satellite.}
\plal{cltt}
\end{plate}

\begin{table}
\begin{tabular}{lll}
Name & Authors				& Journal Reference \\ 
\hline
ARGO & Masi S et al. 1993		& {\it \apjl}   463:L47--L50  \\
ATCA  & Subrahmanyan R et al. 2000	& {\it \mnras}  315:808--822  \\
BAM  & Tucker GS et al. 1997		& {\it \apjl}   475:L73--L76  \\
BIMA  & Dawson KS et al. 2001		& {\it \apjl}   553:L1--L4    \\
BOOM97& Mauskopf PD  et al. 2000	& {\it \apjl}   536:L59--L62  \\
BOOM98& Netterfield CB et al. 2001	& {\it \apj}    In press      \\
CAT99 & Baker JC et al. 1999		& {\it \mnras}  308:1173--1178 \\
CAT96 & Scott PF et al. 1996		& {\it \apjl}   461:L1--L4    \\
CBI   & Padin S et al. 2001		& {\it \apjl}   549:L1--L5    \\
COBE & Hinshaw G, et al. 1996  		& {\it \apj} 	464:L17-L20   \\ 
DASI   & Halverson NW et al. 2001	& {\it \apj}    In press     \\
FIRS & Ganga K, et al. 1994.   		& {\it \apjl} 	432:L15--L18  \\
IAC  & Dicker SR et al. 1999		& {\it \apjl}   309:750--760  \\
IACB & Femenia B, et al. 1998	        & {\it \apj}    498:117--136  \\
QMAP & de Oliveira-Costa A et al. 1998  & {\it \apjl}   509:L77--L80  \\
MAT  & Torbet E et al. 1999		& {\it \apjl}   521:L79--L82  \\
MAX  & Tanaka ST et al. 1996		& {\it \apjl}   468:L81--L84  \\
MAXIMA1& Lee AT et al. 2001		& {\it \apj}    In press     \\
MSAM & Wilson GW et al. 2000		& {\it \apj}    532:57--64    \\
OVRO & Readhead ACS et al. 1989		& {\it \apj}    346:566--587  \\
PYTH & Platt SR  et al. 1997		& {\it \apjl}   475:L1--L4    \\
PYTH5& Coble K  et al. 1999		& {\it \apjl}   519:L5--L8    \\
RING  & Leitch EM  et al. 2000		& {\it \apj}    532:37--56    \\
SASK & Netterfield CB et al. 1997       & {\it \apjl}   477:47--66    \\
SP94 & Gunderson JO, et al. 1995	& {\it \apjl}   443:L57--L60  \\
SP91 & Schuster J et al. 1991		& {\it \apjl}   412:L47--L50  \\
SUZIE & Church SE et al. 1997		& {\it \apj}    484:523--537   \\
TEN  & Guti{\' e}rrez CM, et al. 2000   & {\it \apjl}   529:47--55    \\
TOCO   & Miller AD et al. 1999		& {\it \apjl}   524:L1--L4   \\
VIPER& Peterson JB  et al. 2000		& {\it \apjl}   532:L83--L86  \\
VLA   & Partridge RB et al. 1997	& {\it \apj}    483:38--50    \\
WD    & Tucker GS  et al. 1993		& {\it \apjl}   419:L45--L49   \\
\hline
MAP   & \multicolumn{2}{l} {\tt http://map.nasa.gsfc.gov} \\
Planck& \multicolumn{2}{l} {\tt http://astro.estec.esa.nl/Planck} \\
\hline
\end{tabular}
\caption{CMB experiments shown in \plac{cltt} and references.}
\end{table}

How accurately can the spectra ultimately be measured?
As alluded to above, the fundamental limitation is set by ``cosmic variance'' the fact that
there are only $2\ell + 1$ $m$-samples of the power in each multipole moment.
This leads to an inevitable error of
\begin{eqnarray}
\Delta C_\ell = \sqrt{2 \over 2 \ell +1} C_\ell \,.
\eqnl{deltacl}
\end{eqnarray}
Allowing for
further averaging over $\ell$ in bands of $\Delta \ell \approx \ell$, we
see that the precision in the power spectrum determination scales as $\ell^{-1}$,
i.e.\ $\sim 1\%$ at $\ell=100$ and $\sim 0.1\%$ at $\ell=1000$. 
It is the combination of precision predictions and prospects for precision 
measurements that gives CMB anisotropies their unique stature. 

There are two general caveats to these scalings.  The first is that any source
of noise, instrumental or astrophysical, increases the errors.  If the
noise is also Gaussian and has a known power spectrum, one simply replaces
the power spectrum on the rhs of \eqnc{deltacl} with the sum of the signal
and noise power spectra \citep{Kno95}.  This is the reason that the errors
for the Planck satellite increase near its resolution scale in \plac{cltt} 
(bottom).  Because astrophysical foregrounds are typically 
non-Gaussian it is usually also necessary to remove heavily contaminated
regions, e.g.\ the galaxy.  If the fraction of sky
covered is $f_{\rm sky}$, then the errors increase
by a factor of $f_{\rm sky}^{-1/2}$ and the resulting variance is usually
dubbed ``sample variance'' \citep{ScoSreWhi94}.  
An $f_{\rm sky}=0.65$ was chosen for the Planck satellite.

\subsection{CMB Polarization Field}

While no polarization has yet been detected, general considerations
of Thomson scattering suggest that up to $10\%$ of the anisotropies
at a given scale are polarized.  Experimenters are currently hot on the
trail, with upper limits approaching the expected level
\citep{Hedetal01,Keaetal01}. Thus, we
expect polarization to be an extremely exciting field of study in the
coming decade.

The polarization field can be analyzed in a way
very similar to the temperature field, save for one complication.
In addition to its strength, polarization also has an orientation,
depending on relative strength of two linear polarization states. 
While classical literature has tended to describe polarization locally
in terms of the Stokes parameters $Q$ and $U$\footnote{There is also the
possibility in general of circular polarization, described by Stokes
parameter $V$, but this is absent in cosmological settings.},
recently cosmologists \citep{Sel97,KamKosSte97,ZalSel97} have found that 
the scalar $E$ and pseudo-scalar $B$, linear but non-local 
combinations of $Q$ and $U$, provide a more useful description. 
Postponing the precise definition of $E$ and $B$
until \secc{polarpeaks}, we can, in complete analogy with
\eqnc{thdecompose}, decompose each of them in terms of multipole
moments, and then, following \eqnc{cldef}, consider the power spectra,
\bea
\langle E_{\ell m}^* E_{\ell' m'} \rangle &=& 
\delta_{\ell \ell'}\delta_{m m'} C_{\ell}^{EE}\,, \vs
\langle B_{\ell m}^* B_{\ell' m'} \rangle &=& 
\delta_{\ell \ell'}\delta_{m m'} C_{\ell}^{BB}\,, \vs
\langle \Theta_{\ell m}^* E_{\ell' m'} \rangle &=& 
\delta_{\ell \ell'}\delta_{m m'} C_{\ell}^{\Theta E}\, 
\eqnl{polpow}
.\eea
Parity invariance demands that the cross correlation between the pseudoscalar
$B$ and the scalars $\Theta$ or $E$ vanishes.

The polarization spectra shown in \plac{cltt} [bottom, plotted
in $\mu$K following \eqnc{deltat}] have several notable
features. First, the amplitude of the $EE$ spectrum is indeed down
from the temperature spectrum by a factor of ten. Second, the
oscillatory structure of the $EE$ spectrum is very similar to the
temperature oscillations, only they are apparently out of phase but
correlated with each other. Both of these features are a direct result of the
simple physics of acoustic oscillations as will be shown in 
\secc{acoustic}. The final feature of the polarization
spectra is the comparative smallness of the $BB$ signal. 
Indeed, density perturbations do not produce $B$ modes
to first order.  A detection of substantial $B$ polarization, therefore, would be
momentous. While $E$ polarization effectively doubles our cosmological
information, supplementing that contained in $C_\ell$,
$B$ detection would push us qualitatively forward into new areas
of physics. 

\section{ACOUSTIC PEAKS}
\secl{acoustic}

When the temperature of the Universe was $\sim 3000$K at a redshift $z_*\approx
10^3$, electrons and
protons combined to form neutral hydrogen, an event usually known
as recombination (\citealt{Pee68,ZelKurSun69}; see \citealt{SeaSasSco00} for
recent refinements). 
Before this epoch, free electrons acted as glue between the photons
and the baryons through Thomson and Coulomb scattering,
so the cosmological plasma
was a tightly coupled {\it photon-baryon fluid} \citep{PeeYu70}. 
The spectrum depicted in \plac{cltt} can be explained almost completely
by analyzing the behavior of this pre-recombination fluid. 

In \secc{basics}, we start from the two basic equations of fluid mechanics
and derive the salient characteristics of the anisotropy spectrum:
the existence of peaks and troughs; the spacing between adjacent peaks;
and the location of the first peak. These properties
depend in decreasing order of importance on the initial 
conditions, the energy contents of the Universe before recombination
and those after recombination.  
Ironically, the observational milestones
have been reached in almost the opposite order.
Throughout the 1990's constraints on the location of the first peak
steadily improved culminating with precise determinations from
the TOCO \citep{Miletal99}, Boomerang, \citep{deBetal00} and Maxima-1 
\citep{Hanetal00} experiments (see \plac{cltt} top).  
In the working cosmological model 
it shows up right where it should be if the present energy
density of the Universe is equal to the critical density, i.e.\ if
the Universe is flat.  The skeptic should note that the 
working cosmological model assumes a particular form 
for the initial conditions and energy contents of the Universe before
recombination which we shall see have only recently been tested directly
(with an as yet much lower level of statistical confidence) with the higher peaks.

In \secc{ics} we introduce the initial conditions that apparently
are the source of all clumpiness in the Universe.
In the context of {\it ab initio} models, the term ``initial conditions''
refers to the physical mechanism that generates the primordial small 
perturbations.  In the working cosmological model, this mechanism is
inflation and it sets the initial phase of the oscillations to be the same 
across all Fourier modes.  Remarkably, from this one fact alone comes
the prediction that there will be peaks and troughs in 
the amplitude of the oscillations as a function of wavenumber.  Additionally
the inflationary prediction of an approximately scale-invariant amplitude 
of the initial perturbations implies roughly scale-invariant oscillations in
the power spectrum.  And inflation generically predicts a flat Universe.
These are all falsifiable predictions of the simplest 
inflationary models and they have withstood the test 
against observations to date.

The energy contents of the Universe before recombination
all leave their distinct signatures on the oscillations as discussed 
in \secc{forcing}-\secc{driving}.
In particular, the cold dark matter and baryon signatures have now been seen
in the data \citep{Haletal01,Boom01,Maxima01}. 
The coupling between electrons and photons is not perfect, 
especially as one approaches the epoch
of recombination. 
As discussed in \secc{damping}, 
this imperfect coupling leads to damping in the anisotropy
spectrum: very small scale inhomogeneities are smoothed out. 
The damping phenomenon has now been observed by the CBI experiment
\citep{Padetal01}.  
Importantly, fluid imperfections also generate linear polarization 
as covered in \secc{polarpeaks}.
Because the imperfection is minimal and appears only at small scales, the 
polarization generated is small and has not been detected to date.

After recombination the photons basically travel freely to us today, so the problem
of translating the acoustic inhomogeneities in the photon 
distribution at recombination to
the anisotropy spectrum today is simply one of projection. 
This projection depends almost completely on one number, the angular 
diameter distance between us and the surface of last scattering.   
That number depends on the energy contents
of the Universe after recombination through 
the expansion rate.  
The hand waving projection argument of \secc{basics}
is formalized in \secc{integral}, in the process introducing the popular
code used to compute anisotropies, {\sc cmbfast}.
Finally, we discuss the sensitivity of the acoustic peaks to
cosmological parameters in \secc{sensitivity}.

\subsection{Basics}
\secl{basics}

For pedagogical purposes, let us begin with an idealization of
a perfect photon-baryon fluid and neglect the dynamical
effects of gravity and the baryons. 
Perturbations in this perfect 
fluid can be described by a simple continuity
and an Euler equation that encapsulate the basic properties of
acoustic oscillations.

The discussion of acoustic oscillations will take place
exclusively in Fourier space. 
For example,
we decompose the monopole of the temperature field into
\be
\Theta_{\ell=0,m=0}(\vc{x}) = \int {d^3k \over (2\pi)^3} e^{i\vc{k}\cdot \vc{x}}
	\Theta(\vc{k})
\,,\eqnl{fourier}\ee
and omit the subscript $_{00}$ on the Fourier amplitude.
 Since perturbations
are very small, the evolution equations are linear, and different Fourier
modes evolve independently. Therefore, instead of partial differential equations
for a field $\Theta(\vc{x})$, we have ordinary
differential equations for $\Theta(\vc{k})$. In fact, due to rotational
symmetry, all $\Theta(\vc{k})$ for a given $k$ obey the same
equations.
Here and in the following sections, 
we omit the wavenumber argument $k$ where no 
confusion with physical space quantities will arise.

Temperature perturbations in
Fourier space obey 
\begin{equation}
\dot \Theta = -{1 \over 3} kv_{\gamma}\,,
\eqnl{simplecontinuity}
\end{equation}
This equation for the photon temperature $\Theta$, which does indeed look like the familiar continuity equation
in Fourier space (derivatives $\nabla$ become wavenumbers $k$), has
a number of subtleties hidden in it, due to the cosmological setting.
First, the ``time'' derivative here is actually with respect to {\it
conformal time}
$\eta \equiv \int dt/a(t)$. 
Since we are working in units in which the speed of light $c=1$,
$\eta$ is also the maximum comoving distance a particle could have traveled 
since $t=0$. It is often called
the comoving {\it horizon} or more specifically the comoving {\it particle horizon}. 
The physical horizon is $a$ times the comoving horizon.

Second, the photon fluid 
velocity here $v_\gamma$ has been written as a scalar instead of a vector. 
In the early universe, only the velocity component parallel to the wavevector 
$\vc{k}$ is
expected to be important, since they alone have a source in gravity.
Specifically, $\vc{v}_\gamma = -iv_\gamma \hat\vc{k}$.
In terms of the moments introduced in \secc{observables}, $v_\gamma$
represents a dipole moment directed along $\vc{k}$.
The factor of $1/3$ comes about since continuity conserves photon number
not temperature and the number density $n_\gamma \propto T^3$. 
Finally, we emphasize that, for the time being, we are neglecting the effects
of gravity. 

The Euler equation for a fluid is an expression of momentum conservation.
The momentum density of the photons is $(\rho_\gamma+p_\gamma)v_\gamma$,
where the photon pressure $p_\gamma=\rho_\gamma/3$.
In the absence of gravity and viscous fluid imperfections, pressure gradients 
$\nabla p_\gamma=\nabla \rho_\gamma/3$ supply the only force. 
Since $\rho_\gamma \propto T^4$, this becomes $4 k\Theta\bar\rho_\gamma/3$ 
in
Fourier space.  The Euler equation then
becomes
\begin{equation}
\dot v_{\gamma} = k\Theta \, .
\eqnl{simpleeuler}
\end{equation}

Differentiating the continuity equation and inserting the Euler equation
yields the most basic
form of the oscillator equation 
\begin{equation}
\ddot\Theta + {c_s^2 k^2 }\Theta = 0 \,,
\eqnl{simpleoscillator}
\end{equation}
where $c_s\equiv \sqrt{\dot p/\dot \rho}=1/\sqrt{3}$ is the sound speed in the (dynamically baryon-free) fluid.
What this equation says is that pressure gradients act as a
restoring force to any initial perturbation in the system which thereafter
oscillate at the speed of sound.  Physically these temperature 
oscillations represent the heating and cooling of a fluid that
is compressed and rarefied by a standing sound or acoustic wave.
This behavior continues until recombination.  Assuming negligible
initial velocity perturbations, we have a temperature distribution at
recombination of
\begin{equation}
\Theta(\eta_*) = \Theta(0) \cos(ks_*)\,,
\eqnl{thcos}
\end{equation}
where 
$s= \int c_s d\eta \approx \eta/\sqrt{3}$ is the distance sound can 
travel by $\eta$, usually called the sound horizon.  Asterisks denote
evaluation at recombination $z_*$.

\begin{figure}[t]
\centerline{\epsfxsize=5in\epsffile{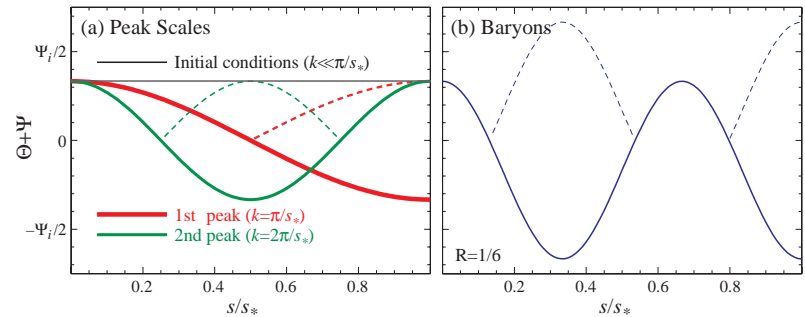}}
\caption{\footnotesize Idealized acoustic oscillations. (a) Peak scales:
the wavemode that completes half an oscillation by recombination sets
the physical scale of the first peak.  Both minima and maxima 
correspond to peaks in power (dashed lines, absolute value) and
so higher peaks are integral multiples of this scale with equal height. 
Plotted here is the idealization of \eqnc{gravityoscillator} 
(constant potentials, no baryon loading).  
(b) Baryon loading.  Baryon loading boosts the amplitudes of every other
oscillation.  Plotted here is the idealization of \eqnc{baryonoscillator} 
(constant potentials and baryon loading $R=1/6$) for the third peak.}
\figl{basic}
\end{figure}

In the limit of scales large compared with the sound horizon $ks \ll 1$,
the perturbation is frozen into its initial conditions.  This is the gist of
the statement that the large-scale anisotropies measured by COBE directly
measure the initial conditions.  On small scales, the amplitude of the 
Fourier modes
will exhibit temporal oscillations, as shown in \figc{basic} 
[with $\Psi=0$, $\Psi_i=3 \Theta(0)$ for this idealization]. 
Modes that are caught at maxima {\it or} minima of their oscillation at
recombination correspond to peaks in the
power, i.e.\ the variance of $\Theta(k,\eta_*)$. Because
sound takes half as long to travel half as far, modes corresponding
to peaks follow a harmonic relationship 
$k_n = n \pi /s_*$, where $n$ is an integer (see \figc{basic}a).

How does this spectrum of inhomogeneities at recombination
appear to us today?
Roughly speaking, a spatial inhomogeneity
in the CMB temperature 
of wavelength $\lambda$ appears as an angular 
anisotropy of scale $\theta \approx \lambda/\da$ where $\da(z)$ is the
comoving angular diameter distance from the observer to redshift
$z$.
We will address this issue more formally in \secc{integral}.
In a flat universe, $\da_* = \eta_0-\eta_* \approx \eta_0$, where
$\eta_0\equiv\eta(z=0)$.  
In harmonic space, the relationship implies a coherent series of
acoustic peaks 
in the anisotropy spectrum, located at
\begin{equation}
\ell_n \approx n \ella, \qquad \ella \equiv \pi \da_*/s_* \,.
\end{equation}

\begin{figure}%
\centerline{\epsfxsize=2.5in\epsffile{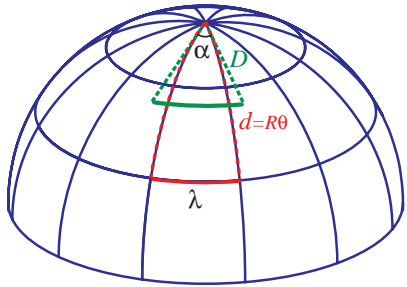}}
\caption{\footnotesize Angular diameter distance.  In a closed universe, objects are 
further than they appear to be from Euclidean (flat) expectations
corresponding to the difference between coordinate distance $d$ and
angular diameter distance $D$.  Consequently, at a fixed coordinate
distance, a given angle corresponds to a smaller spatial scale in 
a closed universe.  Acoustic peaks therefore appear at larger
angles or lower $\ell$ in a closed universe.  The converse is 
true for an open universe. 
}
\figl{geometry}
\end{figure}

To get a feel for where these features should appear, note that in a 
flat matter dominated universe $\eta \propto (1+z)^{-1/2}$ so that
$\eta_*/\eta_0 \approx 1/30 \approx 2^\circ$.  Equivalently $\ell_1 
\approx 200$.  Notice that since we are measuring ratios of distances
the absolute distance scale drops out; we shall see in \secc{driving}
that the Hubble constant sneaks back into the problem because the Universe
is not fully matter-dominated at recombination.

In a spatially curved universe, the angular diameter distance
no longer equals the coordinate distance making the peak locations
sensitive to the spatial curvature of the Universe
\citep{DorZelSun78,KamSpeSug94}.  
Consider first
a closed universe with radius of curvature
$R =  H_{0}^{-1}|\Omega_\tot-1|^{-1/2}$.
Suppressing one spatial coordinate yields
a 2-sphere geometry with the observer situated at the
pole (see \figc{geometry}).  Light travels on
lines of longitude.
A physical scale $\lambda$ at fixed latitude given by
the polar angle $\theta$ subtends an angle
$\alpha = \lambda/R\sin\theta$.
For $\alpha \ll 1$,
a Euclidean analysis would infer a
distance $\da =R\sin\theta$, even though
the {\it coordinate distance} along the arc is
$d = \theta R$; thus
\begin{equation}
\da = R\sin( d / R)\,. 
\end{equation}
For open universes, simply replace $\sin$ with $\sinh$.
The result is that objects in an open (closed) universe are closer
(further) than they appear, as if seen through a lens.  
In fact one way of viewing this effect is as the gravitational
lensing due to the background density (c.f.\ \secc{lensing}).   
A given comoving scale at a fixed distance subtends 
a larger (smaller) angle in
a closed (open) universe than a flat universe.
This strong scaling with spatial curvature indicates that
the observed first peak at $\ell_1 \approx 200$ constrains
the geometry to be nearly spatially flat.   
We will implicitly 
assume spatial flatness in the following sections unless otherwise 
stated.

Finally in a flat dark energy dominated universe, the conformal age of
the Universe decreases approximately as 
$\eta_0 \rightarrow \eta_0 (1 + \ln \Omega_m^{0.085})$.  
For reasonable $\Omega_m$, this causes only a small shift 
of $\ell_1$ to lower multipoles (see \plac{cls}) relative
to the effect of curvature.  Combined with the effect of
the radiation near recombination, 
the peak locations provides a means to measure
the {\it physical} 
age $t_0$ of a flat universe \citep{HuFukZalTeg01}.

\subsection{Initial Conditions}
\secl{ics}

As suggested above, observations of the location of the first peak strongly
point to a flat universe. This is encouraging news for adherents of inflation,
a theory which initially predicted $\Omega_\tot=1$ at a time when few astronomers 
would sign on
to such a high value (see \citealt{LidLyt93} for a review). 
However, the argument
for inflation goes beyond the confirmation of flatness. In particular,
the discussion of the last subsection begs the question: whence $\Theta(0)$,
the initial conditions of the temperature fluctuations?  The answer requires
the inclusion of gravity and considerations of causality which point
to inflation as the origin of structure in the Universe.

The calculations of the typical angular scale of the acoustic oscillations
in the last section are familiar in another context: the horizon problem.
Because the sound speed is near the speed of light, the degree scale also
marks the extent of a causally connected region or particle horizon 
at recombination.  For the picture in the last section to hold, 
the perturbations must have been laid down while the scales in question 
were still far outside the particle horizon\footnote{Recall that the comoving
scale $k$ does not vary with time. At very early times, then, the wavelengh
$k^{-1}$ is much larger than the horizon $\eta$.}.  
The recent observational verification of this basic peak structure presents
a problem potentially more serious than the original 
horizon problem of approximate isotropy: 
the mechanism which smooths fluctuations in the Universe must also 
regenerate them with superhorizon sized correlations at the $10^{-5}$ level.
Inflation is an idea 
that solves both
problems simultaneously.  

The inflationary paradigm postulates that an
early phase of near exponential expansion of the Universe
was driven by a form of energy with negative pressure. In most models, 
this energy is usually provided by the potential 
energy of a scalar field. The inflationary era brings
the observable universe to a nearly smooth and spatially flat state.
Nonetheless, quantum fluctuations in the scalar field are unavoidable and
also carried to large physical scales
by the expansion.  Because an exponential expansion is self-similar in time,
the fluctuations are scale-invariant, i.e.\ in each logarithmic interval in scale
the contribution to the variance of the fluctuations is equal.
Since the scalar field carries the energy density of the
Universe during inflation, its fluctuations induce variations
in the spatial curvature
\citep{GutPi82,Haw82,BarSteTur83}.  
Instead of perfect flatness, inflation 
predicts that each scale will resemble a very slightly open or closed universe.
This fluctuation in the geometry of the Universe is essentially frozen in while the
perturbation is outside the horizon \citep{Bar80}.

Formally, curvature fluctuations are perturbations to the space-space
piece of the metric.  In a Newtonian coordinate system, or gauge, where the metric is 
diagonal, the spatial curvature fluctuation is called $\delta g_{ij} = 
2 a^2 \Phi \delta_{ij}$ (see e.g.\ \citealt{MaBer95}).  
The more familiar Newtonian potential is the time-time fluctuation
$\delta g_{tt} =  2\Psi$ and is 
approximately $\Psi \approx -\Phi$.  Approximate scale invariance then 
says that
$\Delta_\Phi^2 \equiv k^3 P_\Phi(k)/2\pi^2 \propto k^{n-1}$ 
where $P_\Phi(k)$ is the
power spectrum of $\Phi$ and the tilt $n\approx 1$.

Now let us relate the inflationary prediction of scale-invariant curvature fluctuations
to the initial temperature fluctuations. 
Newtonian intuition based on the Poisson equation $k^2\Phi = 4\pi G a^2\delta \rho$ tells us that on
large scales (small $k$) density and hence temperature
fluctuations should be negligible compared with Newtonian potential.
General relativity says otherwise because the Newtonian potential is also 
a time-time fluctuation in the metric.  It corresponds to a temporal shift of
$\delta t/t = \Psi$. The CMB temperature varies as the inverse of the
scale factor, which in turn depends on time as
$a \propto t^{2/[3(1+p/\rho)]}$. Therefore, the fractional change in the
CMB temperature
\be
\Theta = - {\delta a\over a} = -{2\over 3} \left(1+{p\over \rho}\right)^{-1} {\delta t \over
t}\,.
\ee
Thus,  a temporal shift produces a temperature perturbation of
$-\Psi/2$ in the radiation dominated
era (when $p=\rho/3)$ and $-2\Psi/3$ in the matter dominated epoch
($p=0$) (\citealt{Pea91}; 
\citealt{WhiHu97}).  
The initial temperature perturbation is
therefore inextricably linked with the initial gravitational potential
perturbation.  
Inflation predicts scale-invariant initial 
fluctuations in both the CMB temperature and the spatial curvature
in the Newtonian gauge.
 
Alternate models which seek to obey the causality can generate curvature fluctuations only
inside the particle horizon.  Because the perturbations are then
not generated at the same epoch independent of scale, there is no longer a unique
relationship between the phase of the oscillators.  That is, the argument
of the cosine in \eqnc{thcos} becomes $ks_*+\phi(\vc{k})$, where $\phi$ is a phase
which can in principle be different for different wavevectors, 
even those with the same magnitude
$k$. This can lead to
temporal incoherence in the oscillations and 
hence a washing out of the acoustic peaks
\citep{AlbCouFerMag96},
most notably in cosmological defect models \citep{Alletal97,SelPenTur97}.
Complete incoherence is not a strict requirement of causality since there
are other ways to synch up the oscillations.  For example, many
isocurvature models, where the initial spatial curvature is unperturbed, are coherent since their oscillations
begin with the generation of curvature fluctuations at horizon
crossing \citep{HuWhi96a}.  Still they typically have 
$\phi \ne 0$ (c.f.\ \citealt{Tur96}).  
Independent of the angular diameter distance $\da_*$, the
ratio of the peak locations gives the phase:
$\ell_1: \ell_2: \ell_3 \sim 1:2:3$ for $\phi=0$.  
Likewise independent of a constant phase, the spacing of the
peaks $\ell_n - \ell_{n-1} = \ell_A$ gives a measure of the
angular diameter distance \citep{HuWhi96a}.
The observations, which indicate coherent oscillations with $\phi=0$, 
therefore have provided a non-trivial test of 
the inflationary paradigm and supplied 
a substantially more stringent version of the horizon problem
for contenders to solve. 

\subsection{Gravitational Forcing}
\secl{forcing}

We saw above that fluctuations in a scalar field during inflation
get turned into temperature fluctuations via the intermediary of gravity.
Gravity affects $\Theta$ in more ways than this.
The Newtonian potential and spatial curvature alter
the acoustic oscillations by providing a gravitational force on
the oscillator.  The Euler equation (\ref{eqn:simpleeuler})
gains a term on the rhs due to the gradient of
the potential $k\Psi$.  The main effect of gravity then is
to make the oscillations a competition between pressure
gradients $k\Theta$ and potential gradients $k\Psi$ with
an equilibrium when $\Theta+\Psi=0$.  

Gravity also changes the continuity equation.  Since the Newtonian
curvature is essentially a perturbation to the scale factor, changes
in its value also generate temperature perturbations by analogy to
the cosmological redshift $\delta \Theta  = -\delta\Phi$ and so
the continuity equation (\ref{eqn:simplecontinuity}) gains a contribution
of $-\dot\Phi$ on the rhs. 

These two effects bring the oscillator equation (\ref{eqn:simpleoscillator})
to
\begin{equation}
\ddot\Theta + {c_s ^2 k^2 }\Theta = -{k^2 \over 3} \Psi - \ddot \Phi \,.
\end{equation}
In a flat universe and in the absence
of pressure, $\Phi$ and $\Psi$ are constant. Also, in the absence of baryons,
$c_s^2=1/3$ so the new oscillator equation
is identical to \eqnc{simpleoscillator} with $\Theta$ replaced by
$\Theta+\Psi$. The solution 
in the matter dominated epoch is then 
\begin{eqnarray}
[\Theta+\Psi](\eta) &=& [\Theta+\Psi](\eta_{\rm md})\cos(ks)\nonumber\\
                    &=& {1 \over 3}\Psi(\eta_{\rm md})\cos(ks)\,.
\eqnl{gravityoscillator}
\end{eqnarray} 
where $\eta_{\rm md}$ represents the start of the matter dominated epoch
(see \figc{basic}a).
We have used the matter dominated ``initial conditions'' for $\Theta$
given in the previous section assuming large scales, $k s_{\rm md} \ll 1$.

The results from the idealization of 
\secc{basics} carry through with a few exceptions.  Even without
an initial temperature fluctuation to displace the oscillator, acoustic
oscillations would arise by the infall and compression 
of the fluid into gravitational
potential wells.  Since it is 
the {\it effective temperature} $\Theta+\Psi$ that oscillates, they occur even if
$\Theta(0)=0$.  The quantity $\Theta+\Psi$ can be thought of as an
effective temperature in another way: after recombination, photons
must climb out of the potential well to the observer and thus suffer
a gravitational redshift of $\Delta T/T = \Psi$.  The effective temperature
fluctuation is therefore also the observed temperature fluctuation.
We now see that the large scale limit of 
\eqnc{gravityoscillator} recovers the famous Sachs-Wolfe result
that the observed temperature perturbation is $\Psi/3$ 
and overdense regions correspond to cold spots on the sky \citep{SacWol67}.
When $\Psi<0$, although $\Theta$ is positive, the effective temperature
$\Theta+\Psi$ is negative.
The plasma begins effectively rarefied in gravitational potential wells.
As gravity compresses the fluid and pressure resists, rarefaction becomes 
compression and rarefaction again.
The first peak corresponds to the mode that is caught in its first compression
by recombination.  The second peak at roughly half the wavelength corresponds
to the mode that went through a full cycle of compression and rarefaction
by recombination.  We will use this language of the compression 
and rarefaction phase inside initially overdense regions but one should bear
in mind that there are an equal number of initially underdense regions 
with the opposite phase.

\subsection{Baryon Loading}
\secl{loading}

So far we have been neglecting the baryons in the dynamics of
the acoustic oscillations.  To see whether this is a reasonable
approximation consider the photon-baryon momentum density
ratio
$R= (p_b + \rho_b)/(p_\gamma + \rho_\gamma) \approx 
30\Omega_b h^2 (z/10^3)^{-1}$.  For typical values of
the baryon density this number is of order unity at recombination
and so we expect baryonic effects to begin appearing in the oscillations
just as they are frozen in.
 
Baryons are conceptually easy to include in the evolution equations
since their momentum
density provides extra inertia in the joint Euler equation
for pressure and potential gradients to overcome.  Since inertial
and gravitational mass are equal, all terms in the Euler equation save 
the pressure gradient are multiplied 
by $1+R$ leading to the revised oscillator equation \citep{HuSug95a}
\begin{eqnarray}
c_s^2 {d \over d\eta}(c_s^{-2} \dot\Theta)
+ c_s^2 k^2 \Theta &=& -{k^2 \over 3}\Psi  
- c_s^2 {d \over d\eta}(c_s^{-2} \dot\Phi)\,,
\eqnl{baryonoscillator}
\end{eqnarray}
where we have used the fact that the sound speed is reduced
by the baryons to $c_s=1/\sqrt{3(1+R)}$.

To get a feel for the implications of the baryons take the limit
of constant $R$, $\Phi$ and $\Psi$.  Then $d^2(R\Psi)/d\eta^2
(=0)$ may be added to the left hand side to again put the oscillator equation
in the form of \eqnc{simpleoscillator} with $\Theta\rightarrow \Theta + (1+R)\Psi$.
The solution then becomes
\begin{eqnarray}
[\Theta+(1+R)\Psi](\eta) &=& [\Theta+(1+R)\Psi](\eta_{\rm md})\cos(ks) \,.
\eqnl{baryonoscillations}
\end{eqnarray} 
Aside from the lowering of the sound speed which decreases the sound horizon,
baryons have two distinguishing effects:
they enhance the amplitude of the oscillations and shift the equilibrium
point to $\Theta = -(1+R)\Psi$ (see \figc{basic}b).  
These two effects are intimately related 
and are easy to understand since the equations are exactly those of 
a mass $m=1+R$
on a spring in a constant gravitational field.
For the same initial conditions, increasing the mass causes the oscillator
to fall further in the gravitational field leading to larger oscillations
and a shifted zero point.

The shifting of the zero point of the oscillator has significant 
phenomenological consequences.  Since it is still the effective temperature
$\Theta+\Psi$ that is the observed temperature, the zero point shift
breaks the symmetry of the oscillations.  The baryons 
enhance only the compressional phase, i.e.\
every other peak.  For
the working cosmological model these are the first, third, fifth... 
Physically, the extra gravity provided by the baryons enhance 
compression into potential wells.  

These qualitative results 
remain true in the presence of a time-variable $R$.  An additional
effect arises due to the adiabatic damping of an oscillator with
a time-variable mass.  Since the energy/frequency of an oscillator
is an adiabatic invariant, the amplitude must decay as $(1+R)^{-1/4}$.
This can also be understood by expanding the time derivatives in
\eqnc{baryonoscillator} and identifying the $\dot R \dot \Theta$ term 
as the remnant of the familiar expansion drag on baryon velocities.

\begin{figure}[t]
\centerline{\epsfxsize=5in\epsffile{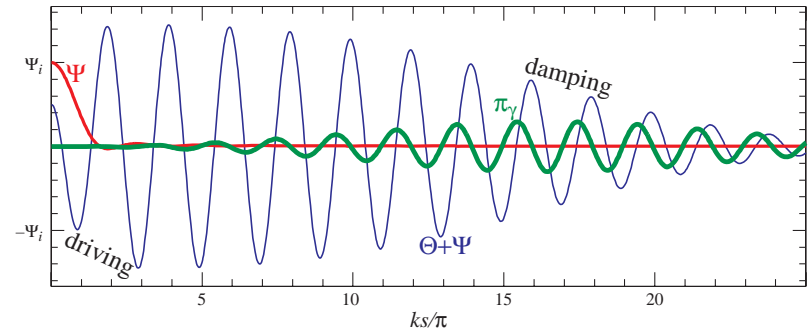}}
\caption{\footnotesize Radiation driving and diffusion damping. The decay of
the potential $\Psi$ drives the oscillator in the radiation dominated
epoch.  Diffusion generates viscosity $\pi_\gamma$, i.e.\ a quadrupole moment
in the temperature, which damps oscillations and generates polarization.
Plotted here is the numerical solution to \eqnc{continuity} 
and \eqnc{Euler} for a mode with wavelength much smaller than 
the sound horizon at decoupling, $ks_* \gg 1$. 
}
\figl{driving}
\end{figure}

\subsection{Radiation Driving}
\secl{driving}

We have hitherto also been neglecting the energy density of the radiation in
comparison to the matter.  The matter-to-radiation ratio scales
as $\rho_m/\rho_r \approx 24 \Omega_m h^2 (z/10^3)^{-1}$
and so is also of order unity at recombination for reasonable parameters.
Moreover fluctuations corresponding to the higher peaks entered the
sound horizon at an earlier time, during radiation domination.

Including the radiation changes the expansion rate of the Universe
and hence the physical scale of the sound horizon at recombination. It
introduces yet another potential ambiguity in the interpretation of
the location of the peaks. 
Fortunately, the matter-radiation ratio has another effect
in the power spectrum by which it can be distinguished.  
Radiation drives the acoustic oscillations
by making the gravitational force evolve with time \citep{HuSug95a}. 
Matter does not.

The exact 
evolution of the potentials is determined by the relativistic Poisson
equation.
But qualitatively, we know that the background density is decreasing with
time, so unless the density fluctuations in the dominant component grow unimpeded
by pressure, potentials will decay.  In particular, in the radiation dominated era
once pressure begins to fight gravity at the first compressional
maxima of the wave, the Newtonian gravitational potential and
spatial curvature must decay (see \figc{driving}).  

This decay actually drives the
oscillations: it is timed to leave the fluid maximally compressed with
no gravitational potential to fight as it turns around.
The net effect is doubled since the redshifting from the spatial metric fluctuation
$\Phi$ also goes away at the same time.
When the Universe becomes matter dominated
the gravitational potential is no longer determined by
photon-baryon density perturbations but by the pressureless
cold dark matter. Therefore, the amplitudes of the acoustic peaks
increase as the cold dark matter-to-radiation ratio decreases
\citep{Sel94,HuSug95a}.
Density perturbations
in any form of radiation will stop growing around horizon crossing
and lead to this effect.  
The net result is that across 
the horizon scale at
matter radiation equality $(k_{\rm eq} \equiv (4-2\sqrt{2})/\eta_{\rm eq})$
the acoustic amplitude increases by a factor of 4-5 \citep{HuSug96}.
By eliminating gravitational potentials, photon-baryon acoustic
oscillations eliminate the alternating peak heights from baryon loading.
The observed high third peak \citep{Haletal01} is a good indication 
that cold dark matter both exists and dominates the energy density 
at recombination.

\subsection{Damping}
\secl{damping}

The photon-baryon fluid has slight imperfections corresponding to
shear viscosity and heat conduction in the fluid \citep{Wei71}.  These imperfections damp
acoustic oscillations.  To consider these effects, we now present
the equations of motion of the system in their full form, including
separate continuity and Euler equations for the baryons.  
Formally the continuity 
and Euler equations follow from the covariant conservation of the
joint stress-energy tensor of the photon-baryon fluid.
Because photon and baryon numbers are separately conserved,
the continuity equations are unchanged, 
\begin{eqnarray}
\dot \Theta = -{k \over 3} v_{\gamma} - \dot\Phi \, , \qquad
\dot \delta_b = -k v_b - 3\dot\Phi \, ,
\label{eqn:continuity}
\end{eqnarray}
where $\delta_b$ and $v_b$ are the density perturbation and
fluid velocity of the baryons.
The Euler equations contain qualitatively new terms 
\begin{eqnarray}
\dot v_{\gamma} &=& k(\Theta + \Psi) - {k \over 6}
\pi_\gamma
        - \dot\tau (v_\gamma - v_b) \, , \nonumber\\
\dot v_b &= &- {\dot a \over a} v_b + k\Psi + \dot\tau(v_{\gamma} - v_b)/R
        \, .
\label{eqn:Euler}
\end{eqnarray}
For the baryons 
the first term on the right accounts for cosmological expansion,
which makes momenta decay as $a^{-1}$. The third term on the right
accounts for momentum exchange in the Thomson scattering between 
photons and electrons
(protons are very tightly coupled to electrons via Coulomb scattering),
with $\dot\tau \equiv n_e\sigma_T a$ the differential Thomson
optical depth, and is compensated by its opposite in the photon Euler equation.
These terms are the origin of heat conduction imperfections.
If the medium is optically thick across a wavelength, $\dot\tau/k\gg 1$ 
and the photons and baryons cannot slip past each other.  
As it becomes optically thin,
slippage dissipates the fluctuations.

In the photon Euler equation there is an extra force on the rhs
due to anisotropic stress gradients or radiation viscosity in the 
fluid, $\pi_\gamma$.  The anisotropic stress is directly proportional to the
quadrupole moment of the photon temperature distribution. 
A quadrupole moment is established by gradients in $v_\gamma$ 
as photons from say neighboring temperature crests meet at
a trough (see \plac{projection}, inset). 
However it is destroyed by scattering.  
Thus $\pi_\gamma = 2 ( k v_\gamma/\dot\tau) A_{\rm v}$, where the order
unity constant can be derived from the Boltzmann equation 
$A_{\rm v} = 16/15$ \citep{Kai83}.  Its evolution is shown in 
\figc{driving}.  With
the continuity \eqnc{simplecontinuity}, $k v_\gamma \approx -3 \dot \Theta$ and
so viscosity takes the form of a damping term.
The heat conduction term can be shown to have a similar effect
by expanding the
Euler equations in $k/\dot \tau$.
The final oscillator equation including both terms becomes
\begin{eqnarray}
c_s^2 {d \over d\eta}(c_s^{-2} \dot\Theta)
+ 
{k^2 c_s^2 \over \dot\tau} [A_{\rm v}+A_{\rm h}]\dot\Theta +
c_s^2 k^2 \Theta &=& -{k^2 \over 3}\Psi  
- c_s^2 {d \over d\eta}(c_s^{-2} \dot\Phi)\,,
\end{eqnarray}
where the heat conduction coefficient $A_{\rm h}= R^2/(1+R)$. 
Thus we expect the
inhomogeneities to be damped by a exponential factor of order $e^{-k^2 \eta/ \dot\tau}$ (see \figc{driving}).
The damping scale $\kd$ is thus of order $\sqrt{\dot\tau/\eta}$,
corresponding to the geometric mean of the horizon and the mean free path.
Damping can be thought of
as the result of the random walk in the baryons that takes photons
from hot regions into cold and vice-versa \citep{Sil68}.  
Detailed numerical integration of the equations
of motion are required to track the rapid growth of the mean free path
and damping length through recombination itself.  
These calculations show that the damping 
scale is of order $\kd s_* \approx 10$ 
leading to a substantial suppression of the oscillations beyond the third peak.  

How does this suppression depend on the cosmological parameters? As the matter
density $\Omega_m h^2$ increases, the horizon $\eta_*$ decreases
since the expansion rate goes up.
Since the diffusion length is proportional to $\sqrt{\eta_*}$, it 
too decreases
as the matter density goes up but not as much as the angular diameter distance 
$\da_*$ which is also inversely proportional to the expansion rate. 
Thus, more matter translates into more damping at a fixed multipole moment; 
conversely, it corresponds to slightly less damping at a fixed peak number. 
The dependence on baryons is controlled by the mean free path which
is in turn controlled by the free electron density: 
the increase in electron density due to an increase in the
baryons is partially offset by a decrease in the ionization fraction due
to recombination.  The net result under the Saha approximation is that the 
damping length scales approximately as $(\Omega_b h^2)^{-1/4}$.  
Accurate fitting formulae for this scale in terms of
cosmological parameters can be found in \citep{HuWhi97d}.  

\subsection{Polarization}
\secl{polarpeaks}

The dissipation of the acoustic oscillations leaves a signature in
the polarization of CMB in its wake (see e.g.\ \citealt{HuWhi97c} and
references therein for a more complete treatment).
Much like reflection off of a surface, Thomson scattering induces
a linear polarization in the scattered radiation.  
Consider incoming radiation in the $-\vc{x}$ direction scattered at right
angles into the $\vc{z}$ direction (see \plac{pol}, left panel).
Heuristically, incoming radiation shakes an electron in the direction
of its electric field vector or polarization $\hat\vc{\epsilon}'$
causing it to radiate with an outgoing polarization parallel to that
direction.  However since the outgoing polarization 
$\hat\vc{\epsilon}$
must be orthogonal to the outgoing direction, incoming radiation that
is polarized parallel to the outgoing direction cannot scatter leaving 
only one polarization state.
More generally, the Thomson differential
cross section $d\sigma_T/d\Omega \propto 
|\hat\vc{\epsilon'} \cdot \hat\vc{\epsilon}|^2.$  

Unlike
the reflection of sunlight off of a surface, the incoming radiation comes
from all angles.  If it were completely isotropic in intensity, 
radiation coming along the $\hat \vc{y}$ would provide the polarization
state that is missing from that coming along $\hat\vc{x}$ 
leaving the net outgoing radiation unpolarized.
Only a quadrupole temperature anisotropy in the radiation
generates a net linear polarization from Thomson scattering.
As we have seen, a quadrupole 
can only be generated causally by the motion of photons and then only
if the Universe is optically thin to Thomson 
scattering across this scale (i.e.\ it is inversely proportional to $\dot\tau$).    
Polarization generation suffers
from a Catch-22: the scattering which generates polarization also
suppresses its quadrupole source. 

The fact that the polarization strength is of order the quadrupole
explains the shape and height of the polarization spectra in \plac{cltt}b.
The monopole and dipole $\Theta$ and $v_\gamma$ are of the same order of magnitude
at recombination, but their oscillations are $\pi/2$ out of 
phase as follows from
\eqnc{simpleoscillator} and \eqnc{thcos}.
Since the quadrupole is of order $kv_\gamma/\dot\tau$ (see \figc{driving}), 
the polarization
spectrum should be smaller than the temperature spectrum by a factor of order $k/\dot\tau$
at recombination. As in the case of the damping, the precise value requires
numerical work \citep{BonEfs87} 
since $\dot\tau$ changes so rapidly near recombination.  Calculations show 
a steady rise in the polarized fraction with increasing $l$ or $k$ 
to a maximum of about ten percent before damping destroys the oscillations and
hence the dipole source. Since $v_\gamma$ is out of
phase with the monopole, the polarization peaks should also be out of phase with the
temperature peaks. Indeed, \plac{cltt}b shows that this is the case.
Furthermore, the phase relation also tells us that the polarization is 
correlated with the temperature perturbations.  The correlation power $C_\ell^{\Theta E}$ 
being the product of the two, exhibits oscillations at twice the acoustic frequency.

Until now, we have focused on the polarization strength without regard 
to its orientation.  The orientation, like a 2 dimensional vector, is
described by two components $E$ and $B$.
The $E$ and $B$ decomposition is simplest to visualize in the small scale
limit, where spherical harmonic analysis coincides with 
Fourier analysis \citep{Sel97}.
Then the wavevector $\vc{k}$ picks out a preferred direction against which
the polarization direction is measured (see \plac{pol}, right panel). 
Since the linear polarization is
a ``headless vector'' that remains unchanged upon a $180^\circ$ rotation, the
two numbers $E$ and $B$ that define it represent polarization
aligned or orthogonal with the wavevector (positive and negative $E$) and
crossed at $\pm 45^\circ$ (positive and negative $B$).  

In linear theory, scalar perturbations like the gravitational 
potential and temperature perturbations have only
one intrinsic direction associated with them, that provided by $\vc{k}$, and
the orientation of the polarization inevitably takes it cue from that
one direction, thereby producing an $E-$mode.  The generalization to
an all-sky characterization of the polarization changes none of these
qualitative features. The $E-$mode and the $B-$mode are formally distinguished
by the orientation of the Hessian of the Stokes parameters which define
the direction of the polarization itself.  This geometric
distinction is preserved under summation of all Fourier modes as
well as the generalization of Fourier analysis to spherical harmonic analysis.

The acoustic peaks in the polarization appear exclusively in the 
$EE$ power spectrum of \eqnc{polpow}.  This distinction is very 
useful as it allows a clean separation of this effect from those 
occuring beyond the scope of the linear perturbation theory 
of scalar fluctuations: in particular, gravitational waves (see 
\secc{GW}) and gravitational lensing (see \secc{lensing}). 
Moreover, in the working cosmological model, the polarization peaks and 
correlation are precise predictions of the temperature peaks as 
they depend on the same physics.   As such their detection 
would represent a sharp test on the implicit assumptions 
of the working model, especially its
initial conditions and ionization history.

\begin{plate}
\centerline{ \epsfxsize = 5.0in \epsfbox{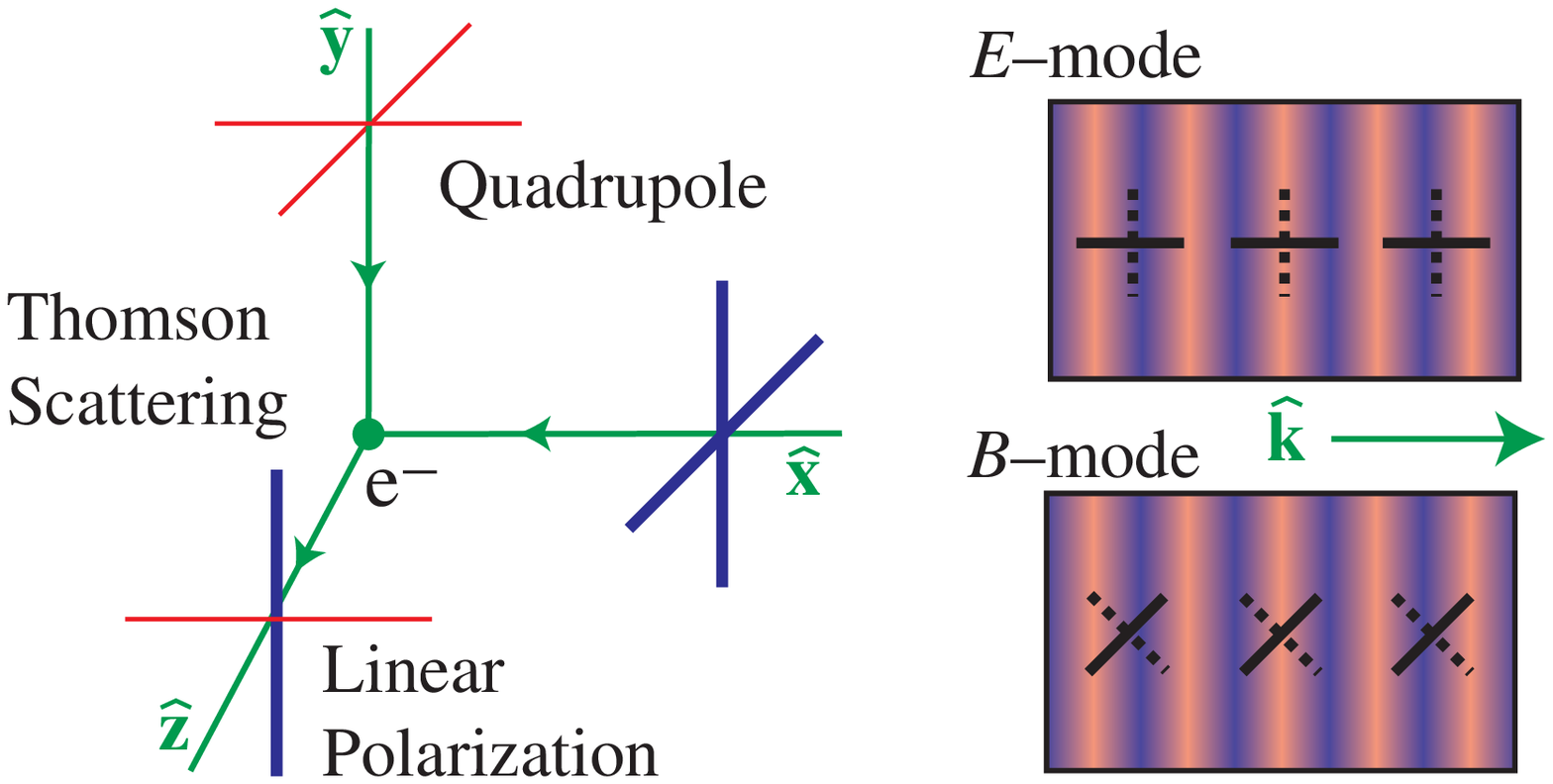} }
\caption{\footnotesize Polarization generation and classification.
Left: Thomson scattering of quadrupole temperature anisotropies (depicted
here in the $\hat x-\hat y$ plane)
generates linear polarization. Right: Polarization in the $\hat x-\hat y$ plane
along the outgoing $\hat z$ axis. The component of the polarization
that is parallel or perpendicular to the wavevector $\vc{k}$ is called the
$E$-mode and the one at $45^\circ$ angles is called the $B$-mode.}
\plal{pol}
\end{plate}

\begin{plate}
\centerline{\epsfxsize=5.5in\epsffile{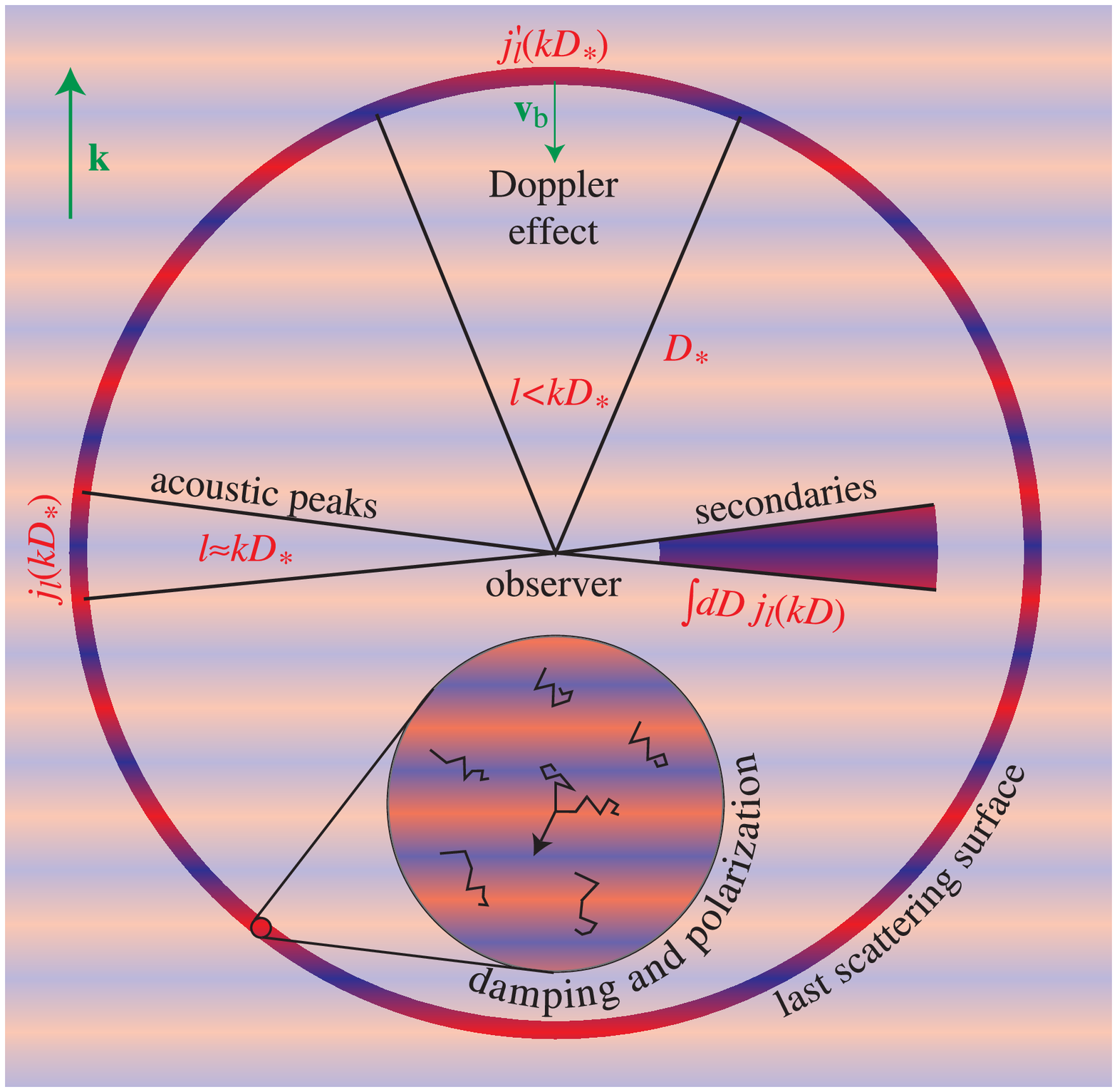}}
\caption{\footnotesize Integral approach.  CMB anisotropies can be thought of as the
line-of-sight projection of various sources of plane wave
temperature and polarization fluctuations: 
the acoustic effective temperature and velocity or Doppler effect
(see \secc{integral}), the quadrupole sources of polarization (see
\secc{polarpeaks}) and secondary sources (see \secc{gsecondaries}, \secc{ssecondaries}).  
Secondary contributions
differ in that the region over which they contribute is thick compared
with the last scattering surface at recombination and the typical wavelength of a perturbation.}
\plal{projection}
\end{plate}

\subsection{Integral Approach}
\secl{integral}

The discussion in the previous sections suffices for a qualitative
understanding of the acoustic peaks in the power spectra of the
temperature and polarization anisotropies.   To refine this treatment
we must consider more carefully the sources of anisotropies and their 
projection into multipole moments.  

Because the description of the acoustic oscillations takes place
in Fourier space, the projection of inhomogeneities at recombination
onto anisotropies today has an added level of complexity. 
An observer today sees the acoustic oscillations
in effective temperature as they appeared on a spherical shell at $\vc{x} =
\da_* \hat n$ at recombination, where $\hat n$ is the direction vector, 
and $\da_* =\eta_0-\eta_*$ is
the distance light can travel between recombination and the present 
(see \plac{projection}).
Having solved for the Fourier amplitude $[\Theta+\Psi](k,\eta_*)$, we can expand the
exponential in \eqnc{fourier} in terms of spherical harmonics, so the
observed anisotropy today is
\be
\Theta({\hat \vc{n}},\eta_0) =  \sum_{\ell m} Y_{\ell m}(\hat \vc{n}) \left[
(-i)^\ell  \int {d^3k\over (2\pi)^3}~ a_\ell(k)
Y^*_{\ell m}(\hat\vc{k}) \right]
,\eqnl{monopole}\ee
where the projected source 
$a_\ell(k)=[\Theta + \Psi](\vc{k},\eta_*)j_\ell(k\da_*)$.
Because the spherical harmonics are orthogonal, \eqnc{thdecompose} implies that
$\Theta_{\ell m}$ today is given by the integral in square brackets today.
A given plane wave actually produces a range of anisotropies in
angular scale as is obvious from \plac{projection}.   The one-to-one mapping
between wavenumber and multipole moment described in \secc{basics}
is only approximately true and comes from the fact that the
spherical Bessel function $j_\ell(k\da_*)$ is strongly peaked 
at $k \da_* \approx \ell$.  Notice that this peak corresponds to
contributions in the direction orthogonal to the wavevector where the correspondence
between $\ell$ and $k$ is one-to-one (see \plac{projection}).

Projection is less straightforward for other sources of anisotropy.
We have hitherto neglected the fact that the acoustic motion of
the photon-baryon fluid also produces a Doppler shift in the radiation
that appears to the observer as a temperature anisotropy as well.
In fact, we argued above that $v_b \approx v_\gamma$ is
of comparable magnitude but out of phase with the effective temperature.
If the Doppler effect projected in the same way as the effective temperature,
it would wash out the acoustic peaks.   However, the Doppler effect has
a directional dependence as well since it is only the line-of-sight
velocity that produces the effect.   Formally, it is a dipole source
of temperature anisotropies and hence has an $\ell=1$ structure.
The coupling of the dipole and plane wave angular momenta
imply that in the projection of the Doppler effect involves a
combination of $j_{\ell\pm 1}$
that may be rewritten as $j_\ell'(x) \equiv dj_\ell(x)/dx$. 
The structure of $j_\ell'$ lacks a strong peak
at $x=\ell$.  Physically this corresponds to the fact that the
velocity is irrotational and hence has no component in the direction
orthogonal to the wavevector (see \plac{projection}).  
Correspondingly, the Doppler effect
cannot produce strong peak structures \citep{HuSug95a}.  
The observed peaks must be acoustic peaks in the 
effective temperature not ``Doppler peaks''.

There is one more subtlety involved when passing from acoustic oscillations
to anisotropies. Recall from \secc{driving} that radiation leads to decay of the
gravitational potentials. Residual radiation after decoupling therefore
implies that the effective temperature is not precisely $[\Theta+\Psi](\eta_*)$.
The photons actually have slightly shallower potentials to climb out of and 
lose
the perturbative analogue of the cosmological redshift, so the
$[\Theta+\Psi](\eta_*)$ overestimates the difference between the true photon temperature
and the observed temperature. This effect of course is already in the 
continuity equation for the monopole \eqnc{continuity} and
so the source in
\eqnc{monopole} gets generalized to
\be
a_\ell(k) = 
\left[\Theta+\Psi\right](\eta_*) j_l(k\da_*)
\, +\, v_b(k,\eta_*) j_\ell'(k\da_*)
\, +\, \int_{\eta_*}^{\eta_0} d\eta (\dot\Psi - \dot\Phi)
j_l(k\da)
\eqnl{isw}
.\ee
The last term vanishes for constant gravitational potentials, but is non-zero
if residual radiation driving exists, as it will in low $\Omega_m h^2$ models.
Note that residual radiation driving is particularly important because it adds
in phase with the monopole: the potentials vary in time only near recombination,
so the Bessel function can be set to $j_l(k\da_*)$ 
and removed from the $\eta$ integral. This complication has the effect of decreasing the multipole value of
the first peak $\ell_1$ as the matter-radiation ratio at recombination decreases \citep{HuSug95a}.
Finally, we mention that time varying potentials can also play a role at very late times
due to non-linearities or the importance of a cosmological constant for example. 
Those contributions, to be discussed more in \secc{ISW}, are sometimes referred
to as late Integrated Sachs-Wolfe effects, and do {\it not} add coherently with
$[\Theta+\Psi](\eta_*)$.

Putting these expressions together and squaring, we obtain the power spectrum
of the acoustic oscillations
\bea
C_\ell &=& {2 \over \pi} \int {dk \over k} k^3 a^2_\ell(k)\,.
\eqnl{clproject}
\eea

This formulation of the anisotropies in terms of projections of
sources with specific local angular structure can be completed to include
all types of sources of temperature and polarization 
anisotropies at any given epoch in time linear or non-linear: 
the monopole, dipole and quadrupole sources arising from density 
perturbations, vorticity and gravitational waves \citep{HuWhi97a}.  
In a curved geometry one replaces the spherical Bessel functions with
ultraspherical Bessel functions \citep{AbbSch86,HuSelWhiZal98}.
Precision in the predictions of the observables
is then limited only by the precision in the prediction of the sources.  
This formulation is ideal for cases where the sources are governed by
non-linear physics even though the CMB responds linearly as we shall see
in \secc{beyond}.

Perhaps more importantly, the widely-used {\sc cmbfast} code 
\citep{SelZal96} exploits these properties 
to calculate the anisotropies in linear perturbation efficiently. 
It numerically solves for 
the smoothly-varying sources on a sparse grid in wavenumber, 
interpolating in the integrals for a handful of $\ell$'s in the
smoothly varying $C_\ell$.  It has largely replaced the original 
ground breaking codes \citep{WilSil81,BonEfs84,VitSil84}
based on tracking the rapid temporal oscillations of the multipole moments
that simply reflect structure in the spherical Bessel
functions themselves.

\begin{plate}
\centerline{\epsfxsize=6in\epsffile{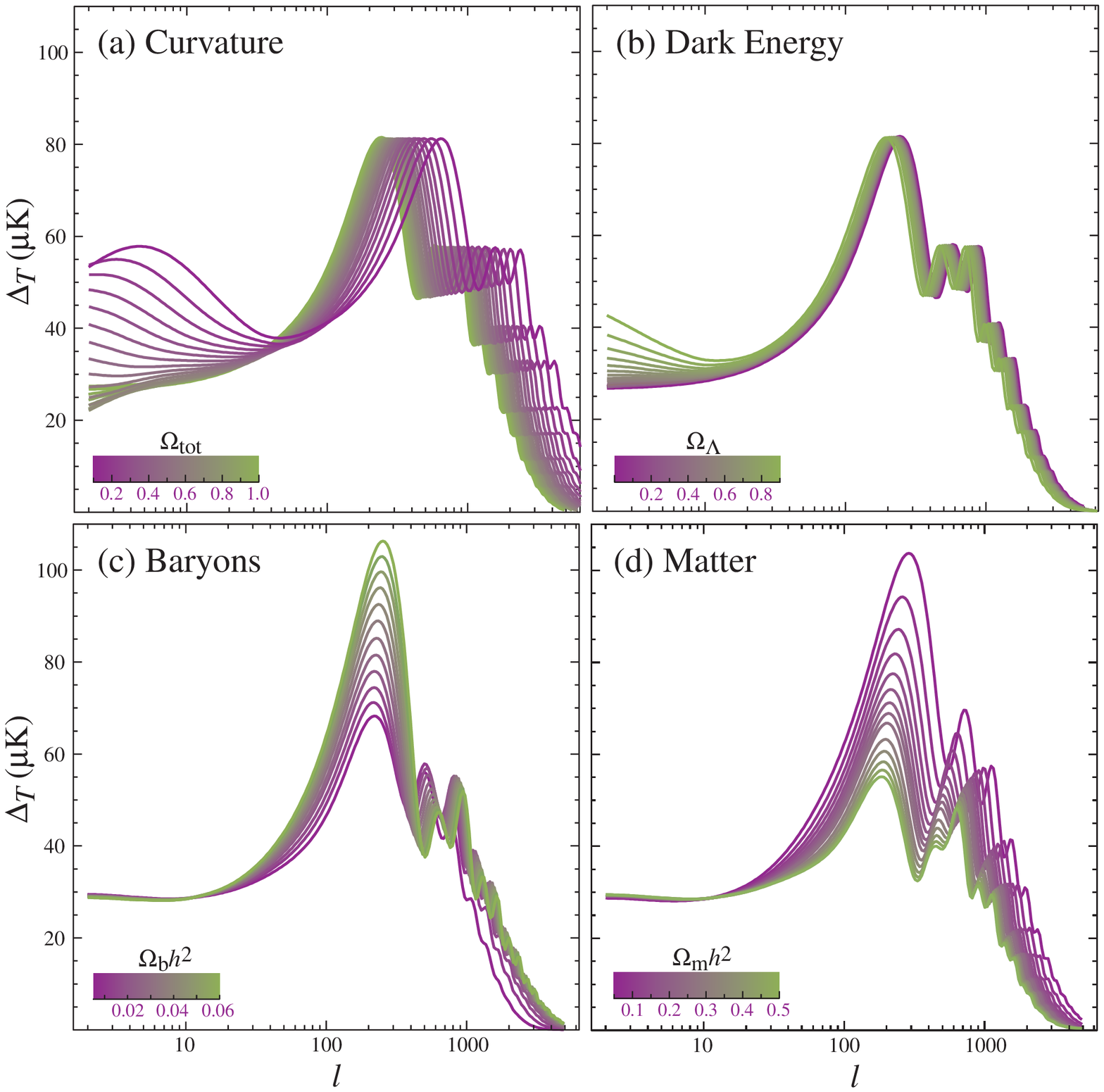}}
\caption{\footnotesize 
Sensitivity of the acoustic temperature spectrum
to four fundamental cosmological parameters (a) the curvature as quantified
by $\Omega_\tot$ (b) the dark energy as quantified by the cosmological
constant $\Omega_\Lambda$ ($w_\Lambda=-1$) (c) the physical baryon density
$\Omega_b h^2$ (d) the physical matter density $\Omega_m h^2$, all
varied around a fiducial model of $\Omega_\tot=1$, $\Omega_\Lambda=0.65$,
$\Omega_b h^2=0.02$, $\Omega_m h^2=0.147$, $n=1$, $z_{\rm ri}=0$,
$E_i=0$.}
\plal{cls}
\end{plate}

\subsection{Parameter Sensitivity}
\secl{sensitivity}

The phenomenology of the acoustic peaks in the temperature and polarization
is essentially described by 
4 observables and the initial conditions \citep{HuSugSil97}.  
These are the angular
extents of the sound horizon $\ella \equiv \pi \da_*/s_*$,
the particle horizon at matter radiation equality 
$\ell_{\rm eq} \equiv k_{\rm eq}\da_*$ and 
the damping scale $\elld
\equiv \kd \da_*$ as well as the value of the baryon-photon momentum
density ratio $R_*$.  
$\ella$ sets the spacing between 
of the peaks; $\ell_{\rm eq}$ and $\elld$ compete to determine their
amplitude through radiation driving and diffusion damping.  
$R_*$ sets the
baryon loading and, along with the potential well depths set by $\ell_{\rm eq}$,
fixes the modulation of the even and odd peak
heights.  
The initial conditions set the phase, or equivalently the location of
the first peak in units of $\ella$, and an overall tilt $n$ 
in the power spectrum.

In the model of \plac{cltt}, these numbers are $\ella = 301$ ($\ell_1=0.73\ella$), 
$\ell_{\rm eq}= 149$, $\elld=1332$, $R_*=0.57$ and $n=1$
and in this family of models 
the parameter sensitivity is approximately \citep{HuFukZalTeg01}
\begin{eqnarray}
    {\Delta \ella \over \ella}
    &\approx&
    -0.24 {\Delta \wm \over \wm}
    +0.07 {\Delta \wb \over \wb}
    -0.17 {\Delta \Omega_{\Lambda} \over \Omega_{\Lambda}}
    -1.1  {\Delta \Omega_\tot \over \Omega_\tot} \,,\nonumber\\
    {\Delta \ell_{\rm eq} \over \ell_{\rm eq}}
    &\approx&
     0.5 {\Delta \wm \over
                           \wm}
    -0.17 {\Delta \Omega_{\Lambda} \over \Omega_{\Lambda}}
    -1.1  {\Delta \Omega_\tot \over \Omega_\tot} \,, \\
{\Delta \elld \over \elld}
    &\approx&
    -0.21 {\Delta \wm \over
                           \wm}
    +0.20 {\Delta \wb \over \wb}
    -0.17 {\Delta \Omega_{\Lambda} \over \Omega_{\Lambda}}
    -1.1  {\Delta \Omega_\tot
    \over \Omega_\tot} \nonumber\,,
\end{eqnarray}
and $\Delta R_*/R_* \approx 1.0 \Delta \wb / \wb$.
Current observations indicate that $\ella = 304\pm 4$, $\ell_{\rm eq}=168
\pm 15$, $\elld=1392 \pm 18$, $R_* = 0.60 \pm 0.06$, and
$n = 0.96 \pm 0.04$ 
(\citealt{KnoCriSko01}; see also 
\citealt{WanTegZal01,Pryetal01,deBetal01}),
if gravitational waves contributions 
are subdominant and the reionization redshift is low as 
assumed in the working cosmological model (see \secc{standard}). 

The acoustic peaks therefore contain three rulers for the angular
diameter distance test for curvature, i.e.\ deviations from $\Omega_\tot=1$.  
However contrary to popular belief, any one of these alone is not
a {\it standard} ruler whose absolute scale is known even in the
working cosmological model.  This is
reflected in the sensitivity of these scales to other cosmological parameters.
For example, the dependence of $\ella$ on $\Omega_m h^2$ and hence the Hubble
constant is quite strong.  But in combination with a measurement of
the matter-radiation ratio from $\elleq$, this degeneracy is broken.

The weaker degeneracy of $\ella$ on the baryons can likewise be broken
from a measurement of the baryon-photon ratio $R_*$.  
The damping scale $\elld$ provides an additional consistency check on the
implicit assumptions in the working model, e.g.\ recombination and the energy contents
of the Universe during this epoch.  What makes the peaks so valuable
for this test is that the rulers are {\it standardizeable} and contain
a built-in consistency check.

There remains a weak but perfect
degeneracy between $\Omega_\tot$ and 
$\Omega_\Lambda$ because they both appear only in $\da_*$.
This is called the angular diameter distance degeneracy in the
literature and can readily be generalized to dark energy components beyond
the cosmological constant assumed here.
Since the effect of $\Omega_\Lambda$ is intrinsically so small, 
it only creates a correspondingly small 
ambiguity in $\Omega_\tot$ for reasonable values of $\Omega_\Lambda$.
The down side is that dark energy can never be isolated through the peaks
alone since it only takes a small amount of curvature to mimic its effects.
The evidence for dark energy through the CMB comes about by allowing
for external information.  The most important is the nearly overwhelming direct
evidence for $\Omega_m < 1$ from local structures in the Universe.  
The second is the measurements of a relatively high Hubble constant $h\approx 0.7$; combined with a relatively low $\Omega_m h^2$ that is
preferred in the CMB data, it implies 
$\Omega_m < 1$ but at low significance currently.

The upshot is that precise measurements of the acoustic peaks
yield precise determinations of four fundamental parameters of the
working cosmological model: $\Omega_b h^2$,
$\Omega_m h^2$, $\da_*$, and $n$.   More generally, the 
first three can be replaced by $\ella$, $\elleq$, $\elld$ and $R_*$
to extend these results to models where the underlying assumptions of the
working model are violated.

\section{BEYOND THE PEAKS}
\secl{beyond}

Once the acoustic peaks in the temperature and polarization power
spectra have been scaled, the days of splendid isolation 
of cosmic microwave background theory, analysis and experiment 
will have ended. 
Beyond and beneath the peaks lies a wealth of
information about the evolution of structure in the Universe and
its origin in the early universe.  As CMB photons traverse the large
scale structure of the Universe on their journey from the recombination 
epoch, they pick up {\it secondary} temperature and polarization anisotropies. 
These depend on the intervening
dark matter, dark energy, baryonic gas density and 
temperature distributions, and even the existence of primordial gravity
waves, so the potential payoff of their detection
is enormous. The price for 
this extended reach is the loss of the ability both to make precise 
predictions,
due to uncertain and/or non-linear physics, and to make precise measurements,
due to the cosmic variance of the primary anisotropies and the
relatively greater importance of galactic and extragalactic
foregrounds.

We begin in \secc{matter} with a discussion of the matter power spectrum
to set the framework for the discussion of secondary anisotropies.  
Secondaries can be divided into two classes: those due to gravitational effects and
those induced by scattering off of electrons. The former are treated in 
\secc{gsecondaries} and the latter in
\secc{ssecondaries}. Secondary anisotropies are often non-Gaussian,
so they show up not only in the power spectra of \secc{observables}, but
in higher point functions as well.
We briefly discuss 
non-Gaussian statistics in \secc{nongaussian}.
All of these topics are subjects
of current research to which this review can only serve as introduction.

\subsection{Matter Power Spectrum}
\secl{matter}

The same balance between pressure and gravity that is responsible for
acoustic oscillations determines the power spectrum of fluctuations in the
non-relativistic matter.  This relationship is often obscured by focussing on
the density fluctuations in the pressureless cold dark matter itself and we
so we will instead consider the matter power spectrum from the perspective
of the Newtonian potential.

\subsubsection{Physical Description}
After recombination, without the pressure of the photons, the baryons
simply fall into the Newtonian potential wells with the cold dark matter, 
an event usually referred to as the end of the Compton drag epoch.   
We claimed in
\secc{driving} that above the horizon at matter-radiation equality
the potentials are nearly constant.  
This follows from the dynamics: where pressure gradients are
negligible,
infall into some initial potential causes a potential flow of 
$v_{\rm tot} \sim (k\eta)\Psi_i$
[see \eqnc{Euler}] and causes density enhancements by continuity of
$\delta_{\rm tot} 
\sim -(k\eta) v_{\rm tot} \sim -(k\eta)^2 \Psi_i$.  The Poisson equation
says that the potential at this later time  $\Psi \sim 
-(k\eta)^{-2}
\delta_{\tot} \sim \Psi_i$ so that this rate of 
growth is exactly right to keep the potential constant.
Formally, this Newtonian argument only applies in general relativity 
for a particular choice of coordinates \citep{Bar80}, but the rule of thumb 
is that if what is driving the expansion (including spatial curvature)
can also cluster unimpeded by pressure, 
the gravitational potential will remain constant.

Because the potential is constant in 
the matter dominated epoch, the large-scale observations of COBE set
the overall amplitude of the potential power spectrum today.  Translated
into density, this is the well-known COBE normalization. It is usually
expressed in terms of $\delta_H$, the matter density perturbation at 
the Hubble scale today.  
Since the observed temperature fluctuation is approximately $\Psi/3$, 
\begin{equation}
{\Delta_T^2 \over T^2} \approx {1 \over 9} \Delta_\Psi^2 \\
	     \approx {1 \over 4} \delta_H^2 \,,
\end{equation}
where the second equality follows from the Poisson equation
in a fully matter dominated universe with $\Omega_m=1$.
The observed COBE fluctuation of $\Delta_T \approx 28\mu$K \citep{Smoetal92}
implies $\delta_H \approx 2 \times 10^{-5}$.  For corrections for
$\Omega_m < 1$ where the potential decays because the dominant driver of
the expansion cannot cluster, see \citet{BunWhi97}. 

On scales below the horizon at matter-radiation equality, 
we have seen in \secc{driving} that pressure gradients from the
acoustic oscillations themselves impede the clustering of the 
dominant component, i.e.\ the photons, and lead to decay in the
potential.  Dark matter density perturbations
remain but grow only logarithmically from their value at horizon crossing, 
which (just as for large scales) is approximately the initial 
potential, $\delta_m \approx -\Psi_i$.  
The potential for modes that have entered the horizon already will 
therefore be suppressed
by $\Psi \propto -\delta_m/k^2 \sim  \Psi_i/k^2$ at matter domination
(neglecting the logarithmic growth)
again according to the Poisson equation. The ratio of $\Psi$ at late times to
its initial
value is called the {\it transfer function}. On large scales, then, the transfer
function is close to one, while it falls off as
 $k^{-2}$ on small scales. 
If the baryons fraction $\rho_b/\rho_m$ is substantial, baryons alter
the transfer function in two ways.  First their inability
to cluster below the sound horizon 
causes further decay in the potential between matter-radiation 
equality and the end of the Compton drag epoch.  Secondly the 
acoustic oscillations in the baryonic velocity field kinematically cause
acoustic wiggles in the transfer function \citep{HuSug96}.  These wiggles 
in the matter power spectrum are related to the acoustic peaks
in the CMB spectrum like twins separated at birth and are actively
being pursued by the largest galaxy surveys \citep{2df}.
For fitting formulae for the transfer function that include these
effects see \citet{EisHu98}.

\subsubsection{Cosmological Implications}
The combination of the COBE normalization, the matter transfer function
and the near scale-invariant initial spectrum of fluctuations
tells us that by the present fluctuations in the cold dark matter 
or baryon density fields will have gone non-linear for all scales 
$k \simgt 10^{-1} h$Mpc$^{-1}$.  
It is a great triumph of the standard cosmological paradigm that
there is just enough growth between $z_* \approx 10^3$ and $z=0$ 
to explain structures in the Universe across a wide range
of scales. 

In particular, since this non-linear scale 
also corresponds to galaxy clusters and measurements of
their abundance yields a robust measure of the power near this
scale for a given matter density $\Omega_m$.  
The agreement between the COBE normalization and the cluster abundance
at low $\Omega_m \sim 0.3-0.4$ and the observed Hubble constant
$h=0.72 \pm 0.08$ \citep{Freetal01}  was pointed out immediately
following the COBE result (e.g.\
\citealt{WhiEfsFre93,BarSil93}) and is one of the
strongest pieces of evidence for the
parameters in the working cosmological model \citep{OstSte95,KraTur95}.

More generally, the comparison between large-scale structure
and the CMB is important in that it breaks degeneracies between
effects due to deviations from power law initial conditions 
and the dynamics of the matter and energy contents 
of the Universe.  Any dynamical effect that
reduces the amplitude of the matter power spectrum corresponds
to a decay in the Newtonian potential that boosts the level
of anisotropy (see \secc{driving} and \secc{ISW}). Massive neutrinos
are a good example of physics that drives the matter power spectrum
down and the CMB spectrum up.

The combination is even more fruitful in the relationship between
the acoustic peaks and the baryon wiggles in the matter power spectrum. 
Our knowledge of the physical distance between adjacent
wiggles provides the ultimate standard candle for cosmology \citep{EisHuTeg99a}.
For example, at very low $z$, the radial distance out to a galaxy is
$cz/H_0$. The unit of distance is therefore $h^{-1}$ Mpc, and a knowledge
of the true physical distance corresponds to a determination of $h$.
At higher redshifts, the radial distance depends sensitively on the
background cosmology (especially the dark energy), so a future measurement
of baryonic wiggles at $z\sim1$ say would be a powerful test of dark energy models.
To a lesser extent,
the shape of the transfer function, which mainly depends on the
matter-radiation scale in $h$ Mpc$^{-1}$, i.e.\ $\Omega_m h$, 
is another standard ruler (see e.g.\ \citealt{TegZalHam01} for 
a recent assessment), more heralded than the wiggles, but less robust
due to degeneracy with other cosmological parameters.

For scales corresponding to $k \simgt 10^{-1} h$ Mpc$^{-1}$, density fluctuations
are non-linear by the present.  Numerical $N$-body simulations show that the dark matter
is bound up in a hierarchy of virialized structures or 
halos (see \citealt{Ber98b} for a review).  
The statistical properties of the dark matter and the dark matter halos 
have been extensively studied in 
the working cosmological model. Less certain are the properties of the baryonic gas.
We shall see that both enter into the consideration of secondary CMB anisotropies.

\subsection{Gravitational Secondaries}
\secl{gsecondaries}

Gravitational secondaries arise from two sources: the differential 
redshift 
from time-variable metric perturbations \citep{SacWol67}
 and gravitational lensing.  There are many examples of the
former, one of which we have already encountered in \secc{integral}
in the context of potential decay in the radiation dominated era.  
Such gravitational potential
effects are usually called the integrated Sachs-Wolfe (ISW) effect in
linear perturbation theory (\secc{ISW}), the Rees-Sciama (\secc{RS})
effect in the non-linear regime, and the gravitational wave effect
for tensor perturbations (\secc{GW}).  
Gravitational waves and lensing also produce $B$-modes
in the polarization (see \secc{polarpeaks}) by which they may be 
distinguished from acoustic polarization.

\subsubsection{ISW Effect} \secl{ISW}
As we have seen
in the previous section, the potential on a given scale 
decays whenever the expansion
is dominated by a component whose effective density is
smooth on that scale.
This occurs at late
times in an $\Omega_m < 1$ model at the end of matter domination and
the onset dark energy (or spatial curvature) domination.
If the potential decays between the time a photon falls into
a potential well and when it climbs out it gets a boost in temperature
of $\delta \Psi$ due to the differential 
gravitational redshift and $-\delta \Phi \approx
\delta \Psi$ due to an accompanying  contraction of the wavelength 
(see \secc{forcing}).  

Potential decay due to residual radiation was introduced in \secc{integral},
but that due to dark energy or curvature at late times induces much different
changes in the anisotropy spectrum. 
What makes the dark energy or curvature contributions
different from those due to radiation 
is the longer length of time over which the potentials decay, 
on order the Hubble time today. Residual radiation produces its
effect quickly, so the distance over which photons feel the effect
is much smaller than the wavelength of the potential
fluctuation. Recall that this meant that $j_l(kD)$ in
the integral in \eqnc{clproject} could be set to $j_l(kD_*)$ and removed
from the integral. The final effect then is proportional to $j_l(kD_*)$
and adds in phase with the monopole.

The ISW projection, indeed the projection of all secondaries,
is much different (see \plac{projection}). Since the duration of 
the potential change is much
longer, photons typically travel through many peaks and troughs of 
the perturbation. This cancellation implies that many modes 
have virtually no impact on the photon
temperature. The only modes which do have an impact are those with wavevectors
perpendicular to the line of sight, so that along the line of sight 
the photon does not pass through crests and troughs. What fraction of
the modes contribute to the effect then? For a given wavenumber $k$ and line
of sight 
instead of the full spherical shell at radius $4\pi k^2 d k$, only the ring 
$2\pi k dk$ with ${\bf \vc{k}} \perp {\bf \vc{n}}$ 
participate.
Thus, the anisotropy induced is suppressed by a 
factor of $k$ (or $\ell=k D$ in angular space).
Mathematically, this arises in the line-of-sight integral of 
\eqnc{clproject} from the integral over the oscillatory 
Bessel function $\int d x j_\ell(x) \approx (\pi/2\ell)^{1/2}$
(see also \plac{projection}).

The ISW effect thus generically
shows up only at the lowest $\ell$'s in the power spectrum (\citealt{KofSta85}).
This spectrum is shown in \plac{secondaries}.  Secondary anisotropy 
predictions in this figure are for a model 
with $\Omega_{\rm tot}=1$, $\Omega_\Lambda=2/3$, $\Omega_b h^2=0.02$, 
$\Omega_m h^2=0.16$, $n=1$, $z_{\rm ri}=7$ and inflationary energy scale $E_i \ll 10^{16}$ GeV.
The ISW effect is especially important in that it is extremely sensitive
to the dark energy: its amount, equation of state and clustering properties 
\citep{CobDodFri97,CalDavSte98,Hu98}.
Unfortunately, being confined to the low multipoles,
the ISW effect suffers severely from the cosmic variance in \eqnc{deltacl} 
in its detectability.  
Perhaps more promising is its correlation with other tracers of
the gravitational potential (e.g.\ X-ray background \citealt{BouCriTur98}
and gravitational lensing, see \secc{lensing}).

\begin{plate}
\centerline{ \epsfxsize = 5in \epsfbox{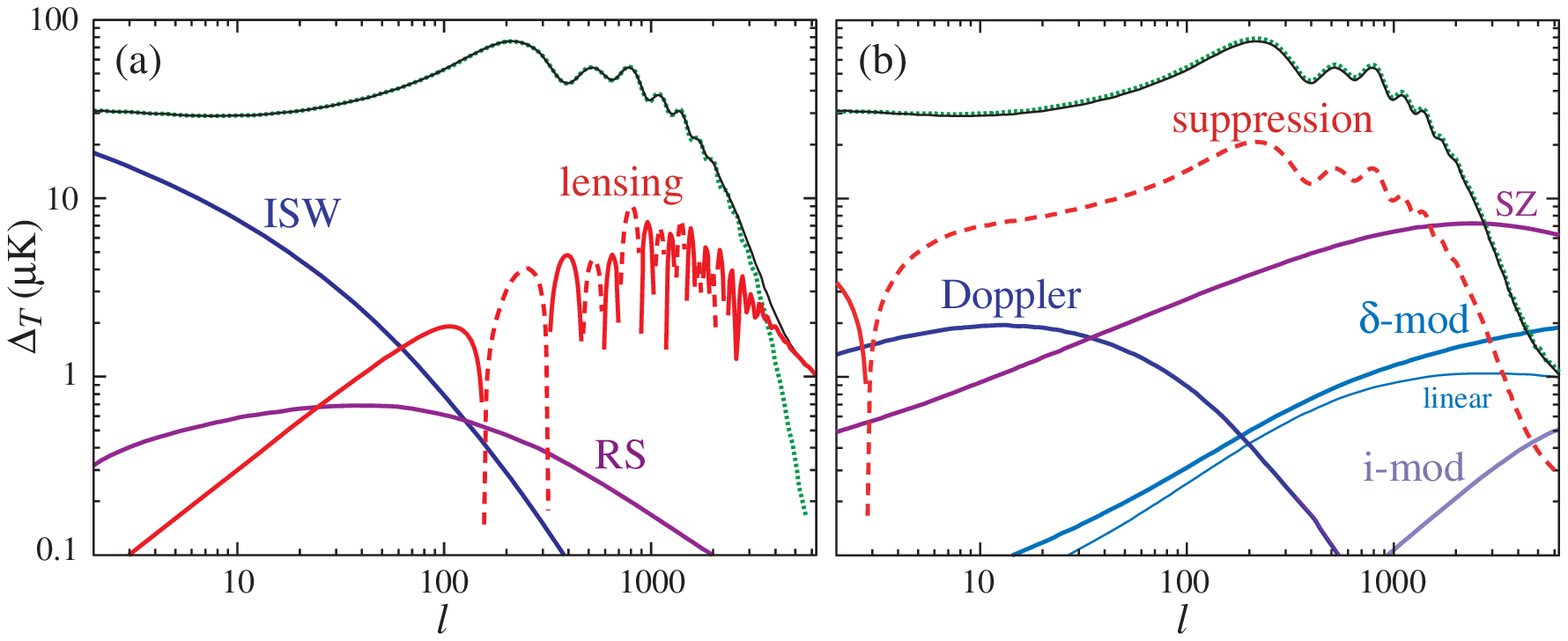} }
\caption{\footnotesize Secondary anisotropies.  (a) Gravitational secondaries:
ISW, lensing and Rees-Sciama (moving halo) effects.
(b) Scattering secondaries: Doppler, density ($\delta$) and 
ionization (i) modulated Doppler, and the SZ 
effects. Curves and model are described in the text.
}
\plal{secondaries}
\end{plate}

This type of cancellation behavior and corresponding suppression of
small scale fluctuations is a common feature of secondary 
temperature and polarization anisotropies from large-scale structure
and is quantified by the Limber
equation \citep{Lim54} and its CMB generalization \citep{HuWhi96a,Hu00b}.
It is the central reason why secondary anisotropies tend to be smaller
than the primary ones from $z_* \approx 10^3$ despite the intervening growth
of structure.

\subsubsection{Rees-Sciama and Moving Halo Effects} \secl{RS}

The ISW effect is linear in the perturbations. 
Cancellation of the ISW effect on small scales leaves second 
order and non-linear analogues in its wake \citep{ReeSci68}.  
From a single isolated structure, the potential along the
line of sight can change not only from evolution in the density profile
but more importantly from its bulk motion across the line of sight.  
In the context of clusters of galaxies, this is called the moving cluster
effect \citep{BirGul83}.   More generally, the bulk motion of 
dark matter halos of all masses contribute to this effect 
\citep{TulLag95,Sel96a}, and their
clustering gives rise to a low level of anisotropies on a range of scales
but is never the leading source of secondary anisotropies on any scale
(see \plac{secondaries}a).

\begin{figure}
\centerline{\epsfxsize=5in\epsffile{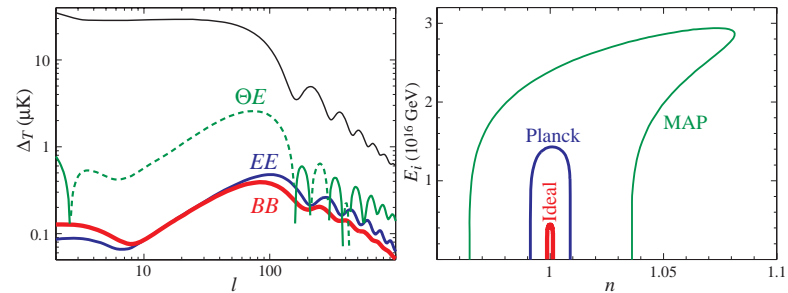}}
\caption{\footnotesize Gravitational waves and the energy scale of
inflation $E_i$. Left: temperature and polarization spectra from an initial
scale invariant gravitational wave spectrum with power 
$\propto E_i^4=(4 \times 10^{16} {\rm GeV})^4$.  Right: 95\% confidence
upper limits statistically achievable on $E_i$ and the scalar tilt $n$
by the MAP and Planck satellites as well as an ideal experiment out
to $\ell=3000$ in the presence of gravitational lensing $B$-modes.}
\figl{tensor}
\end{figure}

\subsubsection{Gravitational Waves} \secl{GW}
A time-variable tensor metric perturbation similarly leaves an imprint in
the temperature anisotropy \citep{SacWol67}. 
A tensor metric perturbation 
can be viewed as a standing gravitational wave and produces a quadrupolar 
distortion in the spatial metric.  If its
amplitude changes, it leaves a quadrupolar distortion in 
the CMB temperature distribution \citep{Pol85}.  
Inflation predicts a nearly scale-invariant spectrum of gravitational waves. 
Their amplitude depends strongly on the
energy scale of inflation,\footnote{$E_i^4 \equiv V(\phi)$, the potential
energy density associated with the scalar field(s) driving inflation.}
(power $\propto E_i^4$
\citealt{RubSazVer82,FabPol83})
and its relationship to the curvature fluctuations discriminates between
particular
models for inflation.  Detection of gravitational waves in the CMB therefore
provides our best hope to study the particle physics of inflation.

Gravitational waves, like scalar fields,
obey the Klein-Gordon equation in a flat universe
and their amplitudes begin oscillating and decaying once the 
perturbation crosses the horizon. While this process occurs 
even before recombination,
rapid Thomson scattering destroys any quadrupole anisotropy that
develops (see \secc{damping}).  This 
fact dicates the general structure of the contributions
to the power spectrum (see \figc{tensor}, left panel): 
they are enhanced at $\ell=2$ the present quadrupole 
and sharply suppressed at multipole larger than that of the first peak
\citep{AbbWis84,Sta85,Crietal93}.  
As is the case for the ISW effect, confinement to the low
multipoles means that the isolation of gravitational waves is severely
limited by cosmic variance.

The signature of gravitational waves in the polarization is
more distinct.  Because gravitational waves 
cause a quadrupole temperature anisotropy  
at the end of recombination, they also generate a polarization.
The quadrupole generated by a gravitational wave has its main angular
variation transverse to the wavevector itself \citep{HuWhi97c}.  
The resulting polarization
that results has components directed both along or orthogonal to 
the wavevector and at 45$^\circ$ degree angles to it.  Gravitational
waves therefore generate a nearly equal amount of $E$ and $B$ mode
polarization when viewed at a distance that is much greater than
a wavelength of the fluctuation  \citep{KamKosSte97,ZalSel97}.
The $B$-component presents a promising means of measuring the gravitational
waves from inflation and hence the energy 
scale of inflation (see \figc{tensor},
right panel). Models of inflation correspond to points in the $n,E_i$
plane \citep{DodKinKol97}. Therefore, the anticipated constraints will discriminate among different
models of inflation, probing fundamental physics at scales well 
beyond those accessible in accelerators.

\subsubsection{Gravitational Lensing} \secl{lensing}
The gravitational potentials of large-scale structure also lens the
CMB photons.  Since lensing conserves surface brightness, it only
affects anisotropies and hence is second order in perturbation
theory \citep{BlaSch87}.  
The photons are deflected according to the angular gradient of
the potential projected along the line of sight with a weighting of
$2 (\da_* - \da)/(\da_* \da)$. Again the cancellation of
parallel modes implies that it is mainly the large-scale potentials
that are responsible for deflections.  
Specifically, the angular gradient
of the projected potential peaks at a multipole $\ell \sim 60$ corresponding
to scales of a $k \sim$ few $10^{-2}$ Mpc$^{-1}$ \citep{Hu00}.  
The deflections are 
therefore coherent below the degree scale.  The coherence of the
deflection should not be
confused with its rms value which in the model of \plac{cltt} has
a value of a few arcminutes.

This large coherence and small amplitude ensures that linear theory in
the potential is sufficient to describe the main effects of
lensing.  Since lensing is a one-to-one mapping of the source
and image planes it simply distorts the images formed from the 
acoustic oscillations in accord with the deflection angle. This
warping naturally also distorts the mapping of physical scales
in the acoustic peaks to angular scales \secc{integral} and hence
smooths features in the temperature and
polarization \citep{Sel96b}.  
The smoothing scale is the coherence scale of
the deflection angle $\Delta \ell \approx 60$ and
is sufficiently wide to alter the acoustic peaks with $\Delta \ell \sim 300$.
The contributions, shown in \plac{secondaries}a are therefore
negative (dashed) on scales corresponding to the peaks.  

For the polarization, the remapping not only smooths the
acoustic power spectrum but actually generates $B$-mode
polarization (see \plac{cltt} and \citealt{ZalSel98b}).  
Remapping by the lenses preserves the orientation of
the polarization but warps its spatial distribution in 
a Gaussian random fashion and hence does not preserve the
symmetry of the original $E$-mode.   The $B$-modes from lensing
sets a detection threshold for gravitational waves for a
finite patch of sky \citep{Hu01c}.

Gravitational lensing also generates a small amount of power in
the anisotropies on its own but this is only noticable beyond
the damping tail where diffusion has
destroyed the primary anisotropies 
(see \plac{secondaries}).  On these small scales, the
anisotropy of the CMB is approximately a pure gradient on the
sky and the inhomogeneous distribution of lenses introduces
ripples in the gradient on the scale of the lenses \citep{SelZal00}.
In fact the moving halo effect of \secc{RS}
can be described as the gravitational
lensing of the dipole anisotropy due to the peculiar motion of
the halo \citep{BirGul83}.

Because the lensed CMB distribution is not linear in the
fluctuations, it
is not completely described by changes in the
power spectrum.  Much of the recent work in the literature
has been devoted to utilizing the non-Gaussianity to isolate
lensing effects \citep{Ber97,Ber98,ZalSel98a,Zal00} and their cross-correlation
with the ISW effect \citep{GolSpe99,SelZal98}.  In particular,
there is a quadratic combination of the anisotropy data that
optimally reconstructs the projected dark matter potentials for
use in this cross-correlation \citep{Hu01b}.   The cross correlation is
especially important in that in a flat universe it is a direct indication
of dark energy and can be used to study the properties of the
dark energy beyond a simple equation of state \citep{Hu01c}. 

\subsection{Scattering Secondaries} 
\secl{ssecondaries}

From the observations both of the lack of 
of a Gunn-Peterson trough  \citep{GunPet65} in quasar spectra and
its preliminary detection \citep{Becetal01}, we know 
that hydrogen was reionized at $z_\ri \simgt 6$.  This is thought to 
occur through the ionizing radiation of the first generation of
massive stars (see e.g.\ \citealt{LoeBar01} for a review). The consequent 
recoupling of CMB photons to the baryons causes a few percent of them
to be rescattered.  Linearly, rescattering induces three changes to the photon
distribution: suppression of primordial anisotropy, generation of large angle
polarization, and a large angle Doppler effect. The latter two are suppressed
on small scales by the cancellation highlighted in \secc{ISW}. Non-linear
effects can counter this suppression; these are the subject of active
research and are outlined in \secc{modulated}.

\subsubsection{Peak Suppression} 
Like scattering before recombination, scattering at late times
suppresses anisotropies in the distribution that
have already formed.  Reionization therefore suppresses the amplitude
of the acoustic peaks by the fraction of photons rescattered,
approximately the optical depth $\sim \tau_{\ri}$ (see \plac{secondaries}b,
dotted line and negative, dashed line, contributions corresponding to
$|\delta \Delta_T^2|^{1/2}$ between the $z_{\rm ri}=7$ and $z_{\rm ri}=0$ models).   
Unlike the plasma before recombination, the medium is optically thin
and so the mean free path and diffusion length of the 
photons is of order the horizon itself. 
New acoustic oscillations cannot form.
On scales approaching
the horizon at reionization, inhomogeneities have yet to be converted 
into anisotropies (see \secc{integral}) and so large angle fluctuations
are not suppressed.  While these effects are relatively large compared
with the expected precision of future experiments, they mimic a 
change in the overall normalization of fluctuations except at the
lowest, cosmic variance limited, multipoles.  

\subsubsection{Large-Angle Polarization}
The rescattered radiation becomes polarized since, as discussed
in \secc{integral} temperature inhomogeneities, 
become anisotropies by projection, passing through quadrupole anisotropies 
when the perturbations are on the horizon scale at any given time.
The result is a bump in the power spectrum of the $E$-polarization on
angular scales corresponding to the horizon at reionization (see \plac{cltt}).  Because 
of the low optical depth of
reionization and the finite range of scales that contribute to the quadrupole,
the polarization contributions are on the order of tenths of $\mu$K on
scales of $\ell \sim $ few.   In a perfect, foreground free world, this
is not beyond the reach of experiments and can be used to isolate
the reionization epoch \citep{HogKaiRes82,ZalSpeSel97}.  
As in the ISW effect, cancellation of contributions along the line of
sight guarantees a sharp suppression of contributions at higher
multipoles in linear theory.  
Spatial modulation of the optical depth due to density and ionization 
(see \secc{modulated}) does produce higher order polarization but
at an entirely negligible level in most models \citep{Hu00b}.

\subsubsection{Doppler Effect}
\secl{doppler}

Naively, velocity fields of order $v \sim 10^{-3}$ 
(see e.g.\ \citealt{StrWil95} for a review) and optical depths of
a few percent would imply a Doppler effect that rivals the acoustic peaks
themselves.  That this is not the case is the joint consequence of the
cancellation described in \secc{ISW} and the 
fact that the acoustic peaks are not ``Doppler peaks'' 
(see \secc{integral}).   Since the Doppler effect comes from the peculiar
velocity along the line of sight, it retains no contributions from linear
modes with wavevectors perpendicular to the line of sight.  But as we
have seen, these
are the only modes that survive cancellation (see \plac{projection} and 
\citealt{Kai84}).  Consequently, the Doppler
effect from reionization is strongly suppressed and is entirely 
negligible below $\ell \sim 10^2$
unless the optical depth in the reionization epoch approaches
unity (see \plac{secondaries}b). 

\subsubsection{Modulated Doppler Effects}
\secl{modulated}

The Doppler effect can survive cancellation if the optical depth has modulations
in a direction orthogonal to the bulk velocity.  
This modulation can be the result of 
either density or ionization fluctuations in the
gas.  Examples of the former include the effect in clusters, 
and linear as well as non-linear large-scale structures.

\smallskip\noindent{\sc Cluster Modulation:}
The strongly non-linear modulation provided by the presence of a galaxy
cluster and its associated gas leads to the {\it kinetic} Sunyaev-Zel'dovich
effect.
Cluster optical
depths on order $10^{-2}$ and peculiar velocities of $10^{-3}$ imply 
signals in the $10^{-5}$ regime in individual arcminute-scale clusters, which 
are of course rare objects.  While this signal is reasonably large,
it is generally dwarfed by the {\it thermal} Sunyaev-Zel'dovich effect
(see \secc{SZ}) and has yet to be detected with high significance 
(see \citealt{Caretal01} and references therein).  The kinetic
 Sunyaev-Zel'dovich effect has negligible impact on 
the power spectrum of anisotropies due to the
rarity of clusters and can be included as part of the
more general density modulation.

\smallskip\noindent{\sc Linear Modulation:}
At the opposite extreme, linear density fluctuations modulate 
the optical depth and give rise to a Doppler effect
as pointed out by \citet{OstVis86} and calculated
by \citet{Vis87} (see also \citealt{EfsBon87}).  
The result is a signal at the $\mu$K level peaking
at $\ell \sim $ few $\times 10^3$ that increases roughly logarithmically
with the reionization redshift (see \plac{secondaries}b).

\smallskip\noindent{\sc General Density Modulation:} 
Both the cluster and linear modulations are limiting cases of the more
general effect of density modulation by the large scale structure of
the Universe.  For the low reionization redshifts currently expected
($z_{\rm ri} \approx 6-7$) most of the effect comes neither from clusters
nor the linear regime but intermediate scale dark matter halos. 
An upper limit to the total effect can be obtained by assuming the 
gas traces the dark matter \citep{Hu00b}
and implies signals on the order of 
$\Delta_T \sim $ few $\mu$K at $\ell > 10^3$ (see \plac{secondaries}b).
Based on simulations, this assumption should hold in the 
outer profiles of
halos \citep{Peaetal01,Lewetal00} but gas pressure will tend to 
smooth out the distribution in the cores
of halos and reduce small scale contributions.  In the absence of 
substantial cooling and star formation, these net effects 
can be modeled under the assumption of hydrostatic equilibrium
\citep{KomSel01} in the halos and included in a halo approach
to the gas distribution \citep{Coo01}.

\smallskip\noindent{\sc Ionization modulation:} 
Finally, optical depth modulation can also come from variations
in the ionization fraction
\citep{AghDesPugGis96,GruHu98,KnoScoDod98}.  
Predictions for this effect are the most
uncertain as it involves both the formation of the first ionizing 
objects and the subsequent radiative transfer of the ionizing radiation
\citep{Bruetal00,Benetal01}.  
It is however unlikely to dominate the density modulated effect except
perhaps at very high multipoles $\ell \sim 10^4$ (crudely estimated,
following \citealt{GruHu98}, in \plac{secondaries}b).

\subsubsection{Sunyaev-Zel'dovich Effect}
\secl{SZ}
Internal motion of the gas in dark matter halos also give rise
to Doppler shifts in the CMB photons.  As in the linear Doppler
effect, shifts that are first order in the velocity are canceled
as photons scatter off of electrons moving in different directions.
At second order in the velocity, there is a residual effect.  For
clusters of galaxies where the temperature of the gas can reach $T_e 
\sim 10$keV,
the thermal motions are a substantial fraction of the speed of light 
$v_{\rm rms} = (3 T_e/ m_e)^{1/2} \sim 0.2$.  The second order effect
represents a net transfer of energy between the hot electron
gas and the cooler CMB and leaves a spectral distortion in the CMB
where photons on the Rayleigh-Jeans side are transferred to the Wien
tail.  This effect is called the thermal Sunyaev-Zel'dovich (SZ) effect
\citep{SunZel72}.
Because the net effect is of order $\tau_{\rm cluster} T_e/m_e 
\propto n_e T_e$, it is a probe of the gas pressure.
Like all CMB effects, once imprinted, distortions
relative to the redshifting background temperature remain unaffected
by cosmological dimming, so one might hope to find clusters at high
redshift using the SZ effect. However, the main effect comes from the 
most massive clusters because of the strong temperature weighting 
and these have formed only recently in the standard cosmological model. 

Great strides have recently been made in observing the SZ effect 
in individual clusters, 
following pioneering attempts that spanned two decades \citep{Bir98}.
The theoretical basis has remained largely unchanged save for
small relativistic corrections as $T_e/m_e$ approches unity.
Both developements are comprehensively reviewed in 
\citep{Caretal01}.
Here we instead consider its implications as a source of secondary 
anisotropies.

The SZ effect from clusters provides the most substantial contribution to
temperature anisotropies beyond the damping tail.  On scales much larger
than an arcminute where clusters are unresolved, contributions to
the power spectrum appear as uncorrelated shot noise ($C_\ell = $ const.
or $\Delta_T \propto \ell$).  The additional contribution due to the
spatial correlation of clusters turns out to be almost negligible in
comparison due to the rarity of clusters \citep{KomKit99}.
Below this scale, contributions turn
over as the clusters become resolved.  Though there has been much
recent progress
in simulations \citep{RefKomEiiSpe00,SelBurPen01,SprWhiHer01} dynamic range
still presents a serious limitation.  

Much recent work has
been devoted to semi-analytic modeling 
following the technique of
\citep{ColKai88}, where the SZ correlations are described in terms
of the pressure profiles of clusters, their abundance and their 
spatial correlations [now commonly referred to an application of 
the ``halo model'' see \citealt{KomKit99,AtrMuc99,Coo01,KomSel01}].  
We show the predictions of a simplified version in \plac{secondaries}b, where
the pressure profile is approximated by the dark matter haloprofile and the 
virial temperature of halo.
While this treatment is comparatively crude, 
the inaccuracies that result are dwarfed by ``missing physics'' in both
the simulations and more sophisticated modelling, e.g.\ the non-gravitational sources and 
sinks of energy that change the temperature and density profile of the cluster,
often modeled as a uniform ``preheating'' of the intercluster 
medium \citep{HolCar01}.   

Although the SZ effect is expected to dominate the power spectrum of secondary
anisotropies, it does not necessarily make the other 
secondaries unmeasurable or
contaminate the acoustic peaks. 
Its distinct frequency signature can be used to isolate it from other
secondaries (see e.g.\ \citealt{CooHuTeg00}).  Additionally, it mainly comes
from massive clusters which are intrinsically rare.  Hence contributions to 
the power spectrum are non-Gaussian and concentrated in rare, spatially 
localized regions. 
Removal of regions identified as clusters through X-rays and optical
surveys  or ultimately high resolution CMB maps themselves  
can greatly reduce contributions at large angular scales where they are
unresolved \citep{PerSpeCen95,KomKit99}.

\subsection{Non-Gaussianity}
\secl{nongaussian}

As we have seen, most of the secondary anisotropies are not linear
in nature and hence produce non-Gaussian signatures.  Non-Gaussianity
in the lensing and SZ signals will be important for their isolation.
The same is true for contaminants such as galactic and extragalactic 
foregrounds.
Finally the lack of an initial non-Gaussianity in the
fluctuations is a testable prediction of the simplest inflationary models 
\citep{GutPi82,BarSteTur83}.  
Consequently, non-Gaussianity
in the CMB is currently a very active field of research.  The  primary
challenge in studies of non-Gaussianity is in choosing the statistic that
quantifies it.  Non-Gaussianity says what the distribution is not, not
what it is.  The secondary challenge is to optimize the statistic
against the Gaussian ``noise'' of the primary anisotropies and instrumental
or astrophysical systematics.

Early theoretical work on the bispectrum, the harmonic analogue of 
the three point function addressed its detectability in the presence
of the cosmic variance of the Gaussian fluctuations \citep{Luo94} and
showed that the inflationary contribution is not expected to be detectable 
in most models (\citealt{AllGriWis87,FalRanSre93}).  
The bispectrum is defined by a triplet of multipoles, or configuration, 
that defines a triangle 
in harmonic space.  The large cosmic variance in an individual configuration
is largely offset by the great number of possible triplets.
Interest was spurred by reports of significant signals in
specific bispectrum configurations in the COBE maps \citep{FerMagGor98} 
that turned out to be due to 
systematic errors \citep{BanZarGor00}.   Recent investigations have
focussed on the signatures of secondary anisotropies 
\citep{GolSpe99,CooHu00a}.  These turn out to be detectable with
experiments that have both high resolution and angular dynamic range
but require the measurement of a wide range of configurations of
the bispectrum.  Data analysis challenges for measuring 
the full bispectrum largely remain to be addressed
(c.f.\ \citealt{Hea98,SpeGol99,PhiKog01}).

The trispectrum, the  harmonic analogue of the four point
function, also has advantages for the study of secondary anisotropies.
Its great number of configurations are specified by a quintuplet of 
multipoles that correspond to the sides and diagonal of a 
quadrilateral in harmonic space \citep{Hu01a}.  The trispectrum is
important in that it quantifies the covariance of the power spectrum
across multipoles that is often very strong in non-linear effects, 
e.g.\ the SZ effect \citep{Coo01}.
It is also intimately related to the power spectra of quadratic combinations
of the temperature field and has been applied to study gravitational
lensing effects \citep{Ber97,Zal00,Hu01a}.

The bispectrum and trispectrum quantify non-Gaussianity in harmonic space,
and have clear applications for secondary anisotropies.  Tests for
non-Gaussianity localized in angular space include the Minkowski functionals
(including the genus) \citep{WinKos97}, the statistics of temperature
extrema \citep{Kogetal96}, and wavelet coefficients \citep{AghFor99}.  
These may be more useful for examining
foreground contamination and trace amounts of topological defects.

\section{DATA ANALYSIS}
\secl{data}

\begin{figure}%
\centerline{\epsfxsize=5in\epsffile{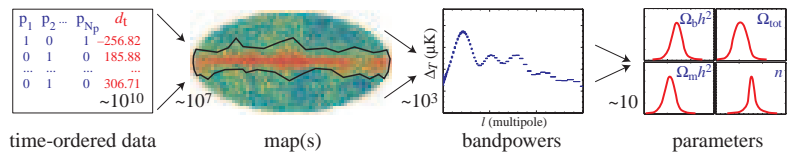}}
\caption{\footnotesize Data pipeline and radical compression.  
Map are constructed for each frequency channel
from the data timestreams, combined and cleaned of foreground contamination 
by spatial (represented here by excising the galaxy) and frequency information.
Bandpowers are extracted from the maps and cosmological parameters from
the bandpowers.  Each step involves a substantial reduction in the
number of parameters needed to describe the data, from potentially
$10^{10} \rightarrow 10$ for the Planck satellite.}
\figl{compress}
\end{figure}

The very large CMB data sets that have begun arriving 
require new, innovative tools of analysis.
The fundamental tool for analyzing CMB data -- the likelihood
function -- has been used since the early days of anisotropy searches
\citep{Read89,Lubin91,DodJub93}. Brute force 
likelihood analyses \citep{TegBun95} were performed even on the relatively large
COBE data set, with six thousand pixels in its map. Present data
sets are a factor of ten larger, and this factor will soon get larger by yet another
factor of a hundred. The brute force approach, the time for which scales as
the number of pixels cubed, no longer suffices.

In response, analysts have devised
a host of techniques that move beyond the early brute force approach. The simplicity
of CMB physics -- due to linearity -- is mirrored in analysis by the apparent
Gaussianity of both the signal and many sources of noise. In the Gaussian limit,
{\it optimal statistics} are easy to identify. These compress the data so that all
of the information is retained, but the subsequent analysis -- because of the
compression -- becomes tractable.

The Gaussianity of the CMB is not
shared by other cosmological systems since gravitational non-linearities
turn an initially Gaussian distribution into a non-Gaussian one. Nontheless, many of the techniques devised to study the CMB
have been proposed for studying: the 3D galaxy distribution \citep{TegStr98}, 
the 2D galaxy distribution \citep{EfsMoo01,HutKn01} the Lyman alpha
forest \citep{SelHui01}, the shear field from weak lensing \citep{HuWhi01},
among others. Indeed, these techniques are now indispensible, powerful tools
for all cosmologists, and we would be remiss not to at least outline them
in a disussion of the CMB, the context in which many of them were developed.

\figc{compress} summarizes the path from the data analyis
starting point, a timestream
of data points, to the end, the determination of cosmological parameters.
Preceding this starting point comes the calibration and the removal
of systematic errors from the raw data, but being experiment specific, 
we do not attempt to cover such 
issues here.\footnote{Aside from COBE, experiments
to date have had a sizable calibration error ($\sim$ 5-10\%) which must be
factored into the interpretation of \plac{cltt}.}
Each step radically compresses the data by reducing the number of
parameters used to describe it.
Although this data pipeline and our discussion below 
are focused on temperature anisotropies,
similar steps have been elucidated 
for polarization \citep{Bun01,TegFam01,LewChaTur01}.

\subsection{Mapmaking}

An experiment can be characterized by the data $d_t$ taken at many different
times; a {\it pointing 
matrix} $P_{ti}$, relating the data timestream to the underlying signal
at pixelized positions indexed by $i$, 
and a noise matrix $C_{d,tt'}$ characterizing the covariance of the noise in the timestream.
A model for the data then is
$d_t = P_{ti}\Theta_i + n_t$ (with implicit sum over the repeating
index $i$); it is the sum of signal plus noise.
Here $n_t$ is drawn from a distribution (often Gaussian) with mean zero
and covariance $\langle n_tn_{t'} \rangle=C_{d,tt'}$. 
In its simplest form the pointing matrix ${\bf P}$ contains rows -- which corresponds to a particular time --
with all zeroes in it except for one column with a one (see \figc{compress}). 
That column corresponds to the particular pixel observed at the time of interest. Typically, a pixel will
be scanned many times during an experiment, so a given column will have many ones
in it, corresponding to the many times the pixel has been observed.

Given this model, a well-posed question is: what is the optimal estimator 
for the signal $\Theta_i$? i.e.\ what is the best way to construct a map?
The answer stems from the likelihood function
$\like$, defined as the probability of getting the data given the theory $\like \equiv
P[{\rm data} | {\rm theory}]$.
In this case, the {\it theory} is the set of parameters $\Theta_i$,
\be
\like_\Theta(d_t)
= {1\over (2\pi)^{N_t/2} \sqrt{\det {\bf C}_{d} }} \exp 
\left[ -{1\over 2}
\left(d_t - P_{ti} \Theta_i\right) C^{-1}_{d,tt'} 
\left(d_{t'} - P_{t'j} \Theta_j\right)\right] \,.
\eqnl{thlike}\ee
That is, the noise, the difference between the data and the
modulated signal, is assumed to be Gaussian with covariance ${\bf C}_d$.

There are two important theorems useful in the construction of a map and 
more generally in each step of the data pipeline \citep{TegTayHea97}.
The first is Bayes' Theorem. In this context, it says that $P[\Theta_i| d_t]$,
the probability that the temperatures are equal to $\Theta_i$ given the
data, is proportional to the likelihood function times a {\it prior}
$P(\Theta_i)$. Thus, with a uniform prior, 
\be
P[\Theta_i| d_t] \propto
P[d_t | \Theta_i] \equiv \like_\Theta(d_t) \,,\ee
with the normalization constant determined by requiring the integral 
of the probability over all $\Theta_i$ to be equal to one. The probability on
the left is the one of interest. The most likely values of $\Theta_i$ therefore
are those which maximize the likelihood function. Since the log of the likelihood
function in question, \eqnc{thlike}, is quadratic in the parameters $\Theta_i$,
it is straightforward to find this maximum point. Differentiating the argument of
the exponential with respect to $\Theta_i$ and setting to zero leads immediately
to the estimator
\be
\hat\Theta_i  = C_{N,ij} P_{jt} C^{-1}_{d,tt'} d_{t'}\,,
\eqnl{map}
\ee
where ${\bf C}_N\equiv ({\bf P}^{\rm tr} {\bf C}_{d}^{-1} {\bf P})^{-1}$.
As the notation suggests, the mean of the estimator is equal to 
the actual $\Theta_i$ and the variance is equal to ${\bf C}_N$. 

The second theorem states that this maximum likelihood estimator is also the
minimum variance estimator.
The {\it Cramer-Rao}
inequality says no estimator can measure the $\Theta_i$ with errors smaller
than the diagonal elements of ${\bf F}^{-1}$,
where the {\it Fisher matrix} is defined as
\be
F_{\Theta,ij} \equiv \left< - {\partial^2 \ln \like_\Theta \over \partial\Theta_i\partial\Theta_j} \right>
.\eqnl{deffish}\ee
Inspection of \eqnc{thlike} shows that, in this case the Fisher matrix is precisely
equal to $C_N^{-1}$. Therefore, the Cramer-Rao theorem implies that the
estimator of \eqnc{map} is optimal: it has the smallest possible variance \citep{Teg97c}. 
No information is lost if the map is used in subsequent analysis instead of the
timestream data, but huge factors of compression have been gained. For example,
in the recent Boomerang experiment \citep{Boom01}, the timestream contained
$2\times 10^8$ numbers, while the map had {\it only} $57,000$ pixels.
The map resulted in compression by a factor of $3500$.

There are numerous complications that must be dealt with in
realistic applications of \eqnc{map}. Perhaps the most difficult is
estimation of ${\bf C}_d$, the timestream 
noise covariance. This typically must be done
from the data itself \citep{FerJaf00,Stompor01}. Even if ${\bf C}_d$ were 
known
perfectly, evaluation of the map involves inverting ${\bf C}_d$, 
a process which 
scales as the number of raw data points cubed. For both of these problems,
the assumed {\it stationarity} of $C_{d,tt'}$ 
(it depends only on $t-t'$) is
of considerable utility.  Iterative techniques to approximate matrix
inversion can also assist in this process \citep{WriHinBen96}.
Another issue which has received
much attention is the choice of pixelization. The community has
converged on the Healpix pixelization scheme\footnote{{\tt
http://www.eso.org/science/healpix/}}, now freely available.

Perhaps the most dangerous complication arises from
astrophysical foregrounds, both within and from outside the
Galaxy, the main ones being synchrotron, bremmsstrahlung, dust
and point source emission. 
All of the main foregrounds have 
different spectral shapes than the blackbody
shape of the CMB. Modern experiments typically 
observe at several different frequencies, so a well-posed
question is: how can we best extract the CMB signal from
the different frequency channels \citep{BouGis99}? The blackbody shape of the
CMB relates the signal in all the channels, leaving one
free parameter. Similarly, if the foreground shapes are
known, each foreground comes with just one free parameter
per pixel. A likelihood function for the data can again
be written down and the best estimator for the CMB amplitude
determined analytically. While in the absence of foregrounds, 
one would extract the CMB signal by weighting the frequency channels
according to inverse noise, when foregrounds are present,
the optimal combination of different frequency
maps is a more clever weighting that subtracts out the foreground contribution
\citep{Dod97}. One can do better if the pixel-to-pixel
correlations of the foregrounds can also be modeled from power spectra
\citep{TegEfs96} or templates derived from external data. 

This picture is complicated somewhat because the
foreground shapes are not precisely known, varying across
the sky, e.g. from
a spatially varying dust temperature.  This too can be modelled 
in the covariance and addressed in the likelihood 
\citep{Teg98,White98}.
The resulting cleaned CMB map is obviously noisier than if foregrounds
were not around, but the multiple channels keep the
degradation managable. For example, the errors on some cosmological
parameters coming from Planck may degrade by almost a
factor of ten as compared with
the no-foreground case. However, many errors will not degrade at
all, and even the degraded parameters will still be
determined with unprecedented precision \citep{Kno99,PruSetBou00,TegEisHu00}. 

Many foregrounds tend to be highly non-Gaussian and in particular
well-localized in particular regions of the map.  These pixels 
can be removed from the map as was done for the region around the
galactic disk for COBE.  This technique can also be highly effective
against point sources.
Indeed, even if there is only
one frequency channel, external foreground templates set
the form of the additional contributions to ${\bf C}_N$, 
which, when properly included, immunize the remaining operations
in the data pipeline to such contaminants \citep{BonJafKno98}.  The
same technique can be used with templates of residual systematics or 
constraints imposed on the data, from e.g.\ the removal of a dipole.

\subsection{Bandpower Estimation}

\figc{compress} indicates that the next step in the compression
process is extracting {\it bandpowers} from the map. What is a 
bandpower and how can it be extracted from the map? To
answer these questions, we must construct a new likelihood function,
one in which the estimated $\Theta_i$ are the data. No theory
predicts an individual $\Theta_i$, but all predict the distribution
from which the individual temperatures are drawn. For example,
if the theory predicts Gaussian
fluctuations, then $\Theta_i$ is distributed as a Gaussian
with mean zero and covariance equal to the sum of the noise covariance
matrix $C_N$ and the covariance due to the finite sample of the 
cosmic signal $C_S$. 
Inverting \eqnc{thdecompose}
and using \eqnc{cldef} for the ensemble average leads to 
\be
C_{S,ij} \equiv \langle \Theta_i \Theta_j \rangle
= \sum_\ell \Delta_{T,\ell}^2 W_{\ell, ij} 
\,,\eqnl{covsig}\ee
where $\Delta_{T,\ell}^2$ depends on the theoretical parameters through
$C_\ell$  (see \eqnc{deltat}). 
Here $W_\ell$, the window function, is proportional to the
Legendre polynomial $P_\ell(\hat n_i\cdot \hat n_j)$ and a beam
and pixel smearing factor $b_\ell^2$. For example, a Gaussian beam of width
$\sigma$, dictates that the observed map is actually a smoothed
picture of true signal, insensitive to structure on scales smaller than
$\sigma$. If the pixel scale is much smaller 
than the beam scale, $b_\ell^2\propto e^{-\ell(\ell+1)\sigma^2}$.
Techniques for handling asymmetric beams have also recently been developed
\citep{Wuetal01,WanGor01,SouRat01}.
Using bandpowers corresponds to assuming that $\Delta_{T,\ell}^2$ is constant
over a finite range, or {\it band}, of $\ell$, equal to $B_a$ for
$\ell_a-\delta \ell_a/2 < \ell < \ell_a + \delta \ell_a/2$. \plac{cltt} gives a sense of the 
width and number of bands $N_b$ probed by existing experiments.

For Gaussian theories, then, the likelihood function is
\be
\like_B (\Theta_i )
= {1\over (2\pi)^{N_p/2} \sqrt{\det {\bf C}_\Theta} } \exp
\left( -{1\over 2} 
\Theta_i C^{-1}_{\Theta,ij} \Theta_j \right)\,,
\ee
where ${\bf C}_\Theta={\bf C}_S+{\bf C}_N$ and $N_p$ is the number of pixels in the map. As before,
$\like_B$ is Gaussian in the anisotropies $\Theta_i$, but in this case $\Theta_i$
are {\it not} the parameters to be determined; the theoretical parameters are
the $B_a$, upon which the covariance matrix depends. Therefore, the likelihood
function is not Gaussian in the parameters, and there is no simple, analytic way to find 
the point in parameter space (which is multi-dimensional depending on the number
of bands being fit) at which $\like_B$ is a maximum. An alternative
is to evaluate $\like_B$ numerically at many points in a grid in parameter
space. The maximum of the $\like_B$ on this grid then determines the
best fit values of the parameters. Confidence levels on say $B_1$ can be
determined by finding the region within which $\int_a^b d B_1 
[ \Pi_{i=2}^{N_b} \int d^B_i ]\,\like_B = 0.95$,
say, for $95\%$ limits. 

This possibility is no longer viable due to the sheer volume of
data. Consider the Boomerang experiment with $N_p=57,000$. A single
evaluation of $\like_B$ involves computation of the inverse and determinant of
the $N_p\times N_p$ matrix ${\bf C}_\Theta$, both of which scale 
as $N_p^3$. While this
single evaluation might be possible with a powerful computer, a single evaluation
does not suffice. The parameter space consists of $N_b=19$ 
bandpowers equally
spaced from $l_a=100$ up to $l_a=1000$. A blindly placed grid on this space would 
require at least ten evaluations in each dimension, so the time required
to adequately evaluate the bandpowers would scale as $10^{19} N_p^3$. No computer
can do this. The situation is rapidly getting worse (better) since Planck will
have of order $10^7$ pixels and be sensitive to of order a $10^3$ bands. 

It is clear that a ``smart'' sampling of the likelihood in parameter space 
is necessary. The numerical problem, searching for the {\it local} maximum
of a function, is well-posed, and a number of search algorithms 
might be used.  $\like_B$ tends to be sufficiently structureless that these
techniques suffice.
\citet{BonJafKno98} proposed the Newton-Raphson method
which has become widely used. One expands the derivative of the
log of the likelihood function -- which vanishes at the true maximum of 
$\like_B$ --
around a trial point in parameter space, $B_a^{(0)}$. Keeping
terms second order in $B_a-B_a^{(0)}$ leads to 
\be
\hat B_a = \hat B_a^{(0)} 
     + \curv_{B,ab}^{-1} {\partial\ln \like_B\over \partial B_b}\,,
\eqnl{lnlike}\ee
where the curvature matrix $\curv_{B,ab}$ is the second derivative of $-\ln\like_B$ with respect to
$B_a$ and $B_b$. Note the subtle distinction between the curvature matrix and
the Fisher matrix in \eqnc{deffish}, ${\bf F} = \langle \hat{\bf F} \rangle$.
In general, the curvature
matrix depends on the data, on the $\Theta_i$. In practice, though, analysts typically use
the inverse of the Fisher matrix in \eqnc{lnlike}. In that case, the estimator becomes
\be
\hat B_a = \hat B_a^{(0)}
+ {1\over 2} F^{-1}_{B,ab} \left(
\Theta_i C_{\Theta,ij}^{-1} {\partial C_{\Theta,jk} \over \partial B_b} C_{\Theta,ki}^{-1} \Theta_i - 
C_{\Theta,ij}^{-1} {\partial C_{\Theta,ji}\over \partial B_b}
\right) \,,
\eqnl{optquad}
\ee
quadratic in the data $\Theta_i$. The Fisher matrix is equal to
\be
F_{B,ab} = {1\over 2} C_{\Theta,ij}^{-1} {\partial C_{\Theta,jk} \over \partial B_a} C_{\Theta,kl}^{-1} 
{\partial C_{\Theta,li} \over \partial B_b}\,.
\eqnl{finfish}
\ee
In the spirit of the Newton-Raphson method, \eqnc{optquad}
is used iteratively but often converges after just a handful of iterations. The usual
approximation is then to take the covariance between the bands
as the inverse of the 
Fisher matrix evaluated at the
convergent point ${\bf C}_{B} = {\bf F}_B^{-1}$. 
Indeed, \citet{Teg97a} derived the identical estimator
by considering all unbiased quadratic estimators, and identifying this one as
the one with the smallest variance. 

Although the estimator in \eqnc{optquad} represents a $\sim 10^{N_b}$ 
improvement
over brute force coverage of the parameter space -- converging in just 
several iterations -- it still requires operations which scale as $N_p^3$.
One means of speeding up the calculations is to transform the data
from the pixel basis to the so-called signal-to-noise basis, based on
an initial guess as to the signal,
and throwing out those modes which have low signal-to-noise 
\citep{Bon95,BunSug95}.  
The drawback is that this procedure
still requires at least one $N_p^3$ operation and potentially many as the
guess at the signal improves by iteration.
Methods to truly avoid this prohibitive $N_p^3$ scaling \citep{OhSpeHin99,WanHan01}
have been devised for experiments with particular scan strategies, but the
general problem remains open. A potentially promising approach involves 
extracting the real space correlation functions as an intermediate step between the
map and the bandpowers \citep{Szaetal01}. Another involves consistently analyzing coarsely
pixelized maps with finely pixelized sub-maps \citep{DorKnoPee01}. 

\subsection{Cosmological Parameter Estimation}

The huge advantage of bandpowers is that they represent the natural
meeting ground of theory and experiment. The above two sections outline
some of the steps involved in extracting them from the observations.
Once they are extracted, any theory can be compared with the observations
without knowledge of experimental details. The simplest way to estimate the cosmological
parameters in a set $c_i$ is to approximate the likelihood as
\begin{equation}
\like_c  (\hat B_a) \approx {1 \over (2\pi)^{N_c/2}  \sqrt{ {\rm det}{\bf C}_{B}} }
\exp\left[ -{1 \over 2} (\hat B_a-B_a) C_{B,ab}^{-1} (\hat B_b-B_b) \right]\,,
\eqnl{likeband}
\end{equation}
and evaluate it at many points in parameter space (the bandpowers depend 
on the cosmological parameters).  
Since the number of cosmological parameters in the working model is $N_c \sim 10$
this represents a final radical compression of information in 
the original timestream which recall has up to $N_t \sim 10^{10}$ data points.

In the approximation that
the band power covariance ${\bf C}_B$ is independent of the parameters $c$, maximizing
the likelihood is the same as minimizing $\chi^2$. 
This has been done by dozens of groups over the last
few years especially since the release of {\sc cmbfast} \citep{SelZal96}, which allows fast computation
of theoretical spectra. Even after all the compression summarized in \figc{compress},
these analyses are still computationally cumbersome due to the large numbers of
parameters varied. Various methods of speeding up spectra computation have been
proposed \citep{TegZal00}, 
based on the understanding of the physics of peaks outlined in \secc{acoustic}, and Monte Carlo explorations of the likelihood function \citep{ChrMeyKnoLue01}.

Again the inverse Fisher matrix gives a quick and dirty estimate of the
errors. Here the analogue of 
\eqnc{deffish} for the cosmological parameters becomes
\begin{equation}
F_{c,ij} = {\partial B_a \over \partial c_i} C^{-1}_{B,ab} {\partial B_b 
\over \partial c_j}\,.
\end{equation}
In fact, this estimate has been widely used to forecast the optimal errors on cosmological
parameters given a proposed experiment and a band covariance matrix $C_B$ which includes diagonal
sample and instrumental noise variance.  The reader should be aware that 
no experiment to date has even come close to achieving the precision 
implied by such a forecast!

As we enter the age of precision cosmology, a number of caveats
will become increasingly important. No theoretical spectra are truly flat in a given band, so the question
of how to weight a theoretical spectrum to obtain $B_a$ can be
important. In principle, one must convolve the theoretical
spectra with window functions \citep{Kno99} distinct from those in \eqnc{covsig}
to produce $B_a$. Among recent experiments,
DASI \citep{Pryetal01} among others have provided these functions. Another complication
arises since the true likelihood function for $B_a$ is not Gaussian, i.e.\ not of the
form in \eqnc{likeband}.  The true distribution is skewed: the cosmic
variance of \eqnc{deltacl} leads to larger errors for an upward fluctuation than for
a downward fluctuation. The true distribution is closer to log-normal \citep{BonJafKno00},
and several groups have already accounted for this in their parameter extractions.

\section{DISCUSSION}

Measurements of the acoustic peaks in the CMB temperature
spectrum have already shown
that the Universe is nearly spatially flat and began with a nearly
scale-invariant spectrum of curvature fluctuations, consistent with the
simplest of inflationary models. In a remarkable confirmation
of a longstanding prediction of Big Bang Nucleosynthesis, the CMB
measurements have now verified that baryons account for about four
percent of the critical density. Further, they suggest that
the matter density is some ten times higher than this, implying the
existence of non-baryonic dark matter and dark energy. 

Future measurements of the morphology of
the peaks in the temperature and polarization
should determine the baryonic and dark matter content of 
the Universe with exquisite precision.  Beyond the peaks, 
gravitational wave imprint on the polarization,
the gravitational lensing of the CMB, 
and gravitational and scattering secondary anisotropies hold the 
promise of understanding the physics of inflation and the impact of
dark energy on structure formation.

The once and future success of the CMB anisotropy enterprise
rests on three equally important pillars: advances in experimental technique, 
precision in theory, and development of 
data analysis techniques.  The remarkable progress in the field over
the last decade owes much to the efforts of researchers in all three
disciplines.  That much more effort will be required to fulfill the bright 
promise of the CMB suggests that the field will remain active and productive
for years to come.

\bigskip
\noindent{\bf ACKNOWLEDGMENTS} 
\medskip

W.H. thanks the hospitality of Fermilab
where this review was written.  W.H. was supported by NASA NAG5-10840 
and the DOE OJI program. S.D. was supported by the DOE, by NASA grant NAG 5-10842 at Fermilab,
by NSF Grant PHY-0079251 at Chicago.

\begin{multicols}{2}

\end{multicols}

\end{document}